\documentclass{article}
\usepackage{arxiv}

\usepackage[utf8]{inputenc} 
\usepackage[T1]{fontenc}    
\usepackage{hyperref}       
\usepackage{url}            
\usepackage{booktabs}       
\usepackage{amsfonts}       
\usepackage{microtype}      
\usepackage{lipsum}
\usepackage{color}
\usepackage{subfigure}
\usepackage{comment}

\usepackage{mathrsfs}
\usepackage{amssymb}
\usepackage{amsmath}
\usepackage{amsthm}
\usepackage{dsfont}

\usepackage{tikz}
\usetikzlibrary{shapes}
\usetikzlibrary{arrows}

\newtheorem{defin}{Definition}

\newtheorem{cor}{Corollary}
\newtheorem{lem}{Lemma}
\newtheorem{prop}{Proposition}

\newtheorem{theorem}{Theorem}

\newtheorem*{rulesparents}{Rules to compute parents}
\newtheorem*{ruleschildren}{Rules to compute children}

\newcommand{\EE}{\mathscr{E}}
\newcommand{\FF}{\mathcal{F}}
\newcommand{\FFF}{\mathscr{F}}
\newcommand{\RR}{\mathscr{R}}
\newcommand{\BB}{\mathds{B}}
\newcommand{\TT}{\mathbb{T}}

\def\JC{\color{red}}

\title{Partial Order on the set of \\Boolean Regulatory Functions}

\author{
	Jos\'e E.R. Cury\\
	Departamento de Automação e Sistemas\\
	Universidade Federal de Santa Catarina, Florianópolis, Brazil\\
	\texttt{jose.cury@ufsc.br}\\
	\And
	Pedro T. Monteiro\\
	INESC-ID, Lisboa, Portugal\\
	Instituto Superior T\'ecnico - Universidade de Lisboa, Lisboa, Portugal\\
	\texttt{Pedro.Tiago.Monteiro@tecnico.ulisboa.pt}\\
	\And
	Claudine Chaouiya\\ 
	Instituto Gulbenkian de Ci\^encia, IGC, Oeiras, Portugal\\
	Aix Marseille Univ, CNRS, Centrale Marseille, I2M, Marseille, France\\
	\texttt{claudine.chaouiya@univ-amu.fr}\\
}

\begin{document}
\maketitle

\begin{abstract}
	Logical models have been successfully used to describe regulatory and signaling networks without requiring quantitative data. However, existing data is insufficient to adequately define a unique model, rendering the parametrization of a given model a difficult task.
	
	Here, we focus on the characterization of the set of Boolean functions compatible with a given regulatory structure, {\it i.e.} the set of all monotone nondegenerate Boolean functions.
	We then propose an original set of rules to locally explore the direct neighboring functions of any function in this set, without explicitly generating the whole set.
	Also, we provide relationships between the regulatory functions and their corresponding dynamics.

	Finally, we illustrate the usefulness of this approach by revisiting Probabilistic Boolean Networks with the model of T helper cell differentiation from Mendoza \& Xenarios.
\end{abstract}



\section{\label{sec:intro}Introduction}
Logical models (Boolean or multi-valued) have been successfully employed to assess dynamical properties of regulatory and signalling networks \cite{abou-jaoude2016}. While the definition of such models does not require quantitative kinetic parameters, it still implies the specification of (logical) regulatory functions to describe the combined effects of regulators upon their targets.
Data on the mechanisms underlying regulatory mechanisms are still scarce, and modellers often rely on generic regulatory functions; for instance, a component is activated if at least one activator is present and no inhibitor is present \cite{mendoza2006}, or if the weighted sum of its regulator activities is above a specific threshold ({\it e.g.,} \cite{Bornholdt:2008aa,Li:2004aa}).

Here, we focus on Boolean gene networks, and we  address two main questions: 1) given a gene network, how complex is the parametrization of a Boolean model consistent with this topology? 2) and  how the choices of regulatory functions impact the dynamical properties of a Boolean model?

To this end, given a gene $g$, we characterize the partially ordered set $\mathcal{F}_g$ of the Boolean regulatory functions compatible with its regulatory structure, {\it i.e.,} with the number and signs of its regulators.
Generically, if a gene $g$ has $n$ regulators, one can in principle define $2^{2^n}$ potential Boolean regulatory functions. This number is then reduced when imposing the functionality of all interactions ({\it i.e.}, all the variables associated with the regulators appear in the function), and a fixed sign of these interactions. We focus on monotone Boolean functions \cite{thieffry1999}, {\it i.e.,} each interaction has a fixed sign (positive or negative).
However, there is no closed expression of the number of monotone Boolean functions on $n$ variables, known as the Dedekind number \cite{korshunov2003,stephen2014}. Actually, it is even unknown for $n>8$. Moreover, even if the functionality constraint further restricts the number of Boolean functions compatible with a given regulatory structure, this number can still be astronomical.  
The set $\mathcal{F}_g$ thus encompasses all monotone, nondegenerate Boolean functions, which can be visualized on a Hasse diagram. In this work, we propose an original algorithm to explore paths in this diagram, that is to determine the local neighboring functions of any Boolean function in the set $\mathcal{F}_g$. 

Section \ref{sec:preliminaries} introduces some preliminaries on sets, partial orders, Boolean functions and Boolean networks. In Section \ref{sec:po_in_f}, we characterize the set $\mathcal{F}_g$ of regulatory functions consistent with the regulatory structure of a given gene $g$. The direct neighbors of any function in $\mathcal{F}_g$ are characterized in Section \ref{sec:neighbors}. Relationships between the regulatory functions and the model dynamics are investigated in Section \ref{sec:impact}. The usefulness of these characterizations is illustrated in Section \ref{sec:applications}, with the consideration of commonly used regulatory functions, and by revisiting Probabilistic Boolean Networks (PBN) as introduced by Shmulevich {\it et al.} \cite{shmulevich2002}. The paper ends with some conclusions and prospects.



\section{\label{sec:preliminaries}Background}

This section introduces some basic concepts and notation that are used in the remainder of the paper.

\subsection{\label{sec:functions}Sets and Partial Orders}

For further detail on the notions introduced here, we refer to relevant text books \cite{caspard2012, davey_priestley}. Given a set $A$, a \emph{Partial Order} on $A$ is a binary relation $\preceq$ on $A$ that satisfies the reflexivity, antisymmetry and transitivity properties.
The pair $(A,\preceq)$ defines a \emph{Partially Ordered Set} (\emph{PO-Set}). A pair of elements $a,b \in A$ is said to be \emph{comparable} in $(A,\preceq)$ if either $(a,b) \in \preceq$ or $(b,a) \in \preceq$. Notation $a\preceq b$ is equally used for $(a,b) \in \preceq$.

A \emph{chain} in a PO-Set $(A,\preceq)$ is a subset of $A$ in which all the elements are pairwise comparable. The symmetrical notion is an \emph{antichain}, defined as a subset of $A$ in which any two elements are incomparable.
Also, an element $a \in A$ is \emph{independent} of an antichain $X\subsetneq A$ if $X \cup \{a\}$ remains an antichain, namely, $a$ is incomparable to every element of $X$. 

A PO-Set $(A,\preceq)$ can be graphically represented by a \emph{Hasse Diagram} ({HD}), in which each element of $A$ is a vertex in the plane and an edge connects a vertex $a\in A$ to a vertex $b\in A$ placed above if $a \preceq b$, and there is no such $c \in A$ such that $a \preceq c \preceq b$ \cite{caspard2012}.

Given a subset $X \subseteq A$, an element $u \in A$ is an \emph{upper bound} of $X$ in the PO-Set $(A,\preceq)$ if $x\preceq u$ for any $x \in X$. Similarly, $l \in A$ is a \emph{lower bound} of $X$ if $l \preceq x$ for any $x \in X$. 
The PO-Set $(A,\preceq)$ is a \emph{Complete Lattice} if any $X \subseteq A$ has a (unique) \emph{least upper bound} and a (unique) \emph{greatest lower bound}.

For a given set $S$, $2^S$ denotes the set of subsets of $S$. A set of elements of $2^S$ whose union contains $S$ is called a \emph{cover} of $S$.


\subsection{\label{sec:boolfunctions}Boolean Functions}

Considering the set $\BB = \{0,1\}$ of the two elements of the Boolean algebra, $\BB^{n}$ denotes the set of Boolean $n$-dimensional vectors $\bold{s} = (s_1,\ldots,s_n)$ with entries in $\BB$.

A Boolean function $f: \BB^{n} \rightarrow \BB$ is \emph{positive} (resp. \emph{negative}) in $s_i$ if $f|_{s_{i}=0} \le f|_{s_{i}=1}$ (resp. $f|_{s_{i}=0} \ge f|_{s_{i}=1}$), where $f|_{s_{i}=0}$ (resp. $f|_{s_{i}=1}$) denotes the value of $f(s_1,\ldots,s_{i-1},0,s_{i+1},\ldots,s_n)$ (resp. $f(s_1,\ldots,s_{i-1},1,s_{i+1},\ldots,s_n)$). We say that $f$ is \emph{monotone} in $s_i$ if it is either positive or negative in $s_i$. $f$ is monotone if it is monotone in $s_i$ for all $i \in \{1,\ldots,n\}$ \cite{crama_hammer}.

Monotone Boolean functions can always be represented in a Disjunctive Normal Form (\emph{DNF}), a disjunction of clauses defined by elementary conjunctions, where each variable appears either in the uncomplemented literal $s_i$ if $f$ is positive in $s_i$, or in the  complemented literal $\neg s_i$ if $f$ is negative in $s_i$. From such a representation, a (unique) canonical representation called the Complete \emph{DNF} of the monotone Boolean function $f$ can be obtained by deleting all redundant clauses, {\it i.e.}, those that are absorbed by other clauses of the original $DNF$ \cite{crama_hammer}.

Determining the number $M(n)$ of monotone Boolean functions for $n$ variables is known as \emph{Dedekind's problem}. This number, also called Dedekind number, is equivalent to the number of antichains in the PO-Set $(2^{\{1,\ldots,n\}},\subseteq)$. $M(n)$ as been computed for values of $n$ up to 8, while asymptotic estimations have been proposed for higher values \cite{caspard2012}. 

A variable $s_i$ is an \emph{essential} variable of a Boolean function $f$ if there is at least one $\bold{s} \in \BB^{n}$ such that $f|_{s_{i}=0} \ne f|_{s_{i}=1}$. A Boolean function is said to be \emph{nondegenerate} if it has no fictitious variables, {\it i.e.}, all variables are \emph{essential} \cite{shmulevich2002}.

Given a Boolean function $f: \BB^{n} \rightarrow \BB$,  $\TT(f)$ denotes the set of vectors $\bold{s} \in \BB^{n}$ for which $f(\bold{s}) = 1$; in other words, $\TT(f)$ is the set of {\it true states} of $f$ \cite{crama_hammer,korshunov2003}.






\subsection{\label{sec:BN}Boolean Networks}

 A \emph{Boolean Network} (BN) is fully defined by a triplet $\RR=(G, R,\mathcal{F})$, where:

\begin{itemize}
	\item $G = \{g_{i}\}_{i = 1, \ldots, n}$ is the set of $n$ regulatory components, each $g_{i}$ being associated with a Boolean variable $s_i$ in $\BB$ that denotes the activity state of $g_i$, {\emph i.e.}, $g_i$ is active (resp. inactive) when $s_i=1$ (resp. $s_i=0$). The set $\BB^{n}$ defines the state space of $\RR$, and $\bold{s} = (s_1, \ldots, s_{n}) \in \BB^{n}$ defines a state of the model;
	\item $R\subseteq G\times G\times\{+,-\}$ is the set of interactions, $(g_i,g_j,+)$ denoting an activatory effect of $g_i$ on $g_j$, and $(g_i,g_j,-)$ an inhibitory effect of $g_i$ on $g_j$;
	\item $\mathcal{F} = \{f_{_i}\}_{i = 1, \ldots, n}$ is the set of Boolean regulatory functions; $f_{i}: \BB^{n} \rightarrow \BB$ defines the target level of component $g_{i}$ for each state $\bold{s} \in \BB^{n}$.
\end{itemize}

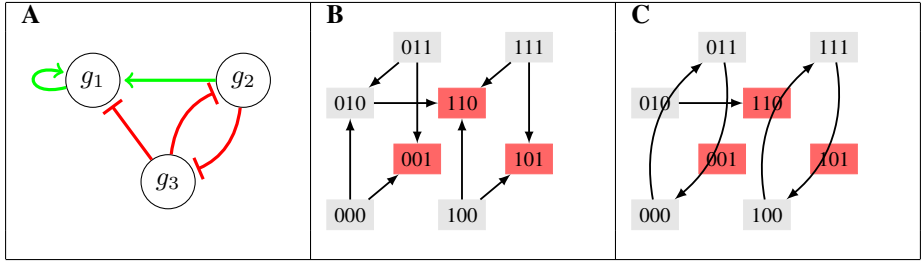
\begin{figure*}[!tpb]
	\centering
\begin{tabular}{|c|c|c|}\hline
\multicolumn{1}{|l|}{\bf A}&\multicolumn{1}{l|}{\bf B}&\multicolumn{1}{l|}{\bf C}\\
\begin{minipage}{0.22\textwidth}
\begin{tikzpicture}[scale=0.5]
	\path(0,0)    node[draw, fill=white!20,shape=circle](G1) {$g_1$};
	\path(4,0)    node[draw, fill=white!20,shape=circle](G2) {$g_2$};
	\path(2,-2.7) node[draw, fill=white!20,shape=circle](G3) {$g_3$};
	\draw[->, thick] (G1) edge[very thick,green, loop left] (G1);
	\draw[->, thick] (G2) edge[very thick,green,shorten >=1.5pt] (G1);
	\draw[-|, thick,shorten >=1.5pt] (G3) edge[very thick,red] (G1);
	\draw[-|, thick,shorten >=1.5pt] (G3) edge[very thick,red, bend left=30] (G2);
	\draw[-|, thick,shorten >=1.5pt] (G2) edge[very thick,red, bend left=30] (G3);
\end{tikzpicture}
\end{minipage}
&\begin{minipage}{0.22\textwidth}\resizebox{0.85\textwidth}{!}{%
\begin{tikzpicture}
	\tikzstyle{stgstate} = [draw,shape=rectangle,font=\small, fill, color=gray!20]
	\tikzstyle{stgedge} = [-latex, thick,font=\sffamily\normalsize\bfseries]

	\def\incx{1.7}
	\def\incy{1.7}
	\colorlet{darkgreen}{green!50!black}

	\node[stgstate] (000) at (0*\incx,0*\incy) {\textcolor{black}{000}};
	\node[stgstate] (100) at (1*\incx,0*\incy) {\textcolor{black}{100}};
	\node[draw,shape=rectangle,font=\small, fill, color=red!60] (001) at (0.6*\incx,0.5*\incy) {\textcolor{black}{001}};
	\node[draw,shape=rectangle,font=\small, fill, color=red!60] (101) at (1.6*\incx,0.5*\incy) {\textcolor{black}{101}};

	\node[stgstate] (010) at (0*\incx,1*\incy) {\textcolor{black}{010}};
	\node[draw,shape=rectangle,font=\small, fill, color=red!60] (110) at (1*\incx,1*\incy) {\textcolor{black}{110}};
	\node[stgstate] (011) at (0.6*\incx,1.5*\incy) {\textcolor{black}{011}};
	\node[stgstate] (111) at (1.6*\incx,1.5*\incy) {\textcolor{black}{111}};

	\draw[stgedge] (010) edge (110);
	\draw[stgedge] (000) edge (010);
	\draw[stgedge] (100) edge (110);
	\draw[stgedge] (011) edge (001);
	\draw[stgedge] (111) edge (101);
	\draw[stgedge] (000) edge (001);
	\draw[stgedge] (100) edge (101);
	\draw[stgedge] (011) edge (010);
	\draw[stgedge] (111) edge (110);
\end{tikzpicture}
}\end{minipage}
&\begin{minipage}{0.22\textwidth}\resizebox{0.85\textwidth}{!}{%
\begin{tikzpicture}
	\tikzstyle{stgstate} = [draw,shape=rectangle,font=\small, fill, color=gray!20]
	\tikzstyle{stgedge} = [-latex, thick,font=\sffamily\normalsize\bfseries]

	\def\incx{1.7}
	\def\incy{1.7}
	\colorlet{darkgreen}{green!50!black}

	\node[stgstate] (000) at (0*\incx,0*\incy) {\textcolor{black}{000}};
	\node[stgstate] (100) at (1*\incx,0*\incy) {\textcolor{black}{100}};
	\node[draw,shape=rectangle,font=\small, fill, color=red!60] (001) at (0.6*\incx,0.5*\incy) {\textcolor{black}{001}};
	\node[draw,shape=rectangle,font=\small, fill, color=red!60] (101) at (1.6*\incx,0.5*\incy) {\textcolor{black}{101}};

	\node[stgstate] (010) at (0*\incx,1*\incy) {\textcolor{black}{010}};
	\node[draw,shape=rectangle,font=\small, fill, color=red!60] (110) at (1*\incx,1*\incy) {\textcolor{black}{110}};
	\node[stgstate] (011) at (0.6*\incx,1.5*\incy) {\textcolor{black}{011}};
	\node[stgstate] (111) at (1.6*\incx,1.5*\incy) {\textcolor{black}{111}};

	\draw[stgedge] (010) edge (110);
	\draw[-latex, thick,font=\sffamily\normalsize\bfseries] (000) to[bend left]  (011);
	\draw[-latex, thick,font=\sffamily\normalsize\bfseries] (011) to[bend left]  (000);
	\draw[stgedge] (100) to[bend left]   (111);
	\draw[stgedge] (111) to[bend left]   (100);
	
\end{tikzpicture}
}\end{minipage}\\&&\\\hline
\end{tabular}
\caption{Example of a Boolean Network with: (A) the regulatory graph, where normal (green) arrows represent activations and hammerhead (red) arrows represent inhibitions; (B) the asynchronous STG, considering the following Boolean regulatory functions $f_{1} = s_1 \vee (s_2 \land \neg s_3)$, $f_{2} = \neg s_3$ and $f_{3} = \neg s_2$; (C) the synchronous STG, for the same regulatory functions. Stable states in both STGs are denoted in red.
}
\label{fig:toyBN}
\end{figure*}

In the corresponding {\em regulatory graph} $(G,R)$, nodes represent regulatory components ({\it e.g.} genes) and directed edges represent signed regulatory interactions (positive for activations and negative for inhibitions).
 Figure~\ref{fig:toyBN}-A shows an example of a regulatory graph with 3 components: a mutual inhibition between $g_2$ and $g_3$, and a self-activation of $g_1$, which is further activated by $g_2$ and repressed by $g_3$. 
 
 The set of the regulators of a component $g_i$ is denoted $G_i\,=\,\{g_j\in G, (g_j,g_i,+)\!\in\! R \mbox{ or } (g_j,g_i,-)\!\in\! R\}$.
Note that the regulatory function of a component $g_i$ may be defined over the states of its regulators (rather than over the states of the full set of components): $\forall g_i\in G,\, f_i:\BB^{|G_i|}\rightarrow \BB$; it thus specifies how regulatory interactions are combined to affect the state of $g_i$. In other words, one can define the regulatory functions over only their essential variables.

The dynamics of a BN is represented as a {\em State Transition Graph} (STG), where each node represents a state, and directed edges represent transitions between states. The STG of a BN $\RR=(G,R,\FF)$ can be formally defined by $\EE = (\BB^{|G|}, \mathcal{T})$, where:
\begin{itemize}
	\item $\BB^{|G|}$ is the state space of $\RR$;
	\item $\mathcal{T} \subseteq \BB^{|G|} \times \BB^{|G|}$ is a transition relation (or transition function), where $(\bold{s},\bold{s}') \in \mathcal{T}$ whenever state $\bold{s}$ is connected to state $\bold{s}'$.
\end{itemize}
Assuming an \emph{asynchronous update mode}, in which components are updated independently \cite{abou-jaoude2016,Le-Novere:2015aa,thomas91a}, we have that
$(\bold{s},\bold{s}') \in \mathcal{T}^{\rm{async}}$ iff:

$$\left\{\begin{array}{l}
\exists i \in \{1,\ldots,n\},\, f_i(\bold{s})=\neg s_i=s^{\prime}_i ,\\
	\forall j \in \{1,\ldots,n\},\, j\neq i,\,f_j(\bold{s})=s^{\prime}_j. \\
\end{array}\right.$$

Hence, if in state $\bold{s}$ several components are such that $f_i(\bold{s})\neq s_i$ ({\it i.e}, are called to update their values), $\bold{s}$ has as many outgoing transitions. 

Under a \emph{synchronous update mode}, all states have at most one outgoing transition, that is $(\bold{s},\bold{s}') \in \mathcal{T}^{\rm{sync}}$ iff:
$$
\forall i\in \{1,\ldots,n\},\,f_i(\bold{s})=s^{\prime}_i. \\
$$

Dynamics are affected by the choice of the (a/synchronous) update except \emph{stable states}, which are states $\bold{s}$ such that $\forall i,\, f_i(\bold{s})=s_i$, are conserved \cite{abou-jaoude2016}, see Figure \ref{fig:toyBN}. Those states are of biological interest as they often correspond to specific phenotypes or cell fates. In the STG, stable states correspond to terminal strongly connected components reduced to a single state. 
Other attractors refer to cyclic or oscillatory behaviors, which are terminal  strongly connected components encompassing multiple states in the STG. While cyclic attractors are also biologically relevant, they may greatly differ depending on the update mode. In contrast to the synchronous update mode, which amounts to consider that underlying mechanisms have exactly the same delays, it is generally acknowledged that the asynchronous update mode is more realistic \cite{abou-jaoude2016,Saadatpour2010,thomas91a}. This is the update we will consider in the reminder of this paper.


\section{\label{sec:po_in_f}Characterizing the set of consistent regulatory functions }

Here, we thus focus on a generic component $g_i$ of a BN, and we show that the set $\mathcal{F}_i$ of the regulatory functions that comply with the regulatory interactions targeting $g_i$ is a {PO-Set}. Properties of this {PO-Set} give an insight on how a particular choice of a function affects the behavior of the sole $g_i$ ({\it i.e.} affect the transitions between states differing on their $i^{th}$ components). Generalization to the complete STG then derives from the combinations of the transition graphs of the individual components.

\begin{figure}[!tpb]
	\centering
\begin{tabular}{|c|c|c|}\hline
\multicolumn{1}{|l|}{\bf A}&\multicolumn{1}{l|}{\bf B}&\multicolumn{1}{l|}{\bf C}\\
\begin{minipage}{0.18\textwidth}\resizebox{\textwidth}{!}{
\begin{tikzpicture}
	\tikzstyle{stgstate} = [draw,shape=rectangle,font=\small, fill, color=gray!20]
	\tikzstyle{stgedge} = [-latex, thick,font=\sffamily\normalsize\bfseries]

	\def\incx{1.7}
	\def\incy{1.7}
	\colorlet{darkgreen}{green!50!black}

	\node[stgstate] (000) at (0*\incx,0*\incy) {\textcolor{black}{000}};
	\node[stgstate] (100) at (1*\incx,0*\incy) {\textcolor{black}{100}};
	\node[draw,shape=rectangle,font=\small, fill, color=red!60] (001) at (0.6*\incx,0.5*\incy) {\textcolor{black}{001}};
	\node[draw,shape=rectangle,font=\small, fill, color=red!60] (101) at (1.6*\incx,0.5*\incy) {\textcolor{black}{101}};

	\node[stgstate] (010) at (0*\incx,1*\incy) {\textcolor{black}{010}};
	\node[draw,shape=rectangle,font=\small, fill, color=red!60] (110) at (1*\incx,1*\incy) {\textcolor{black}{110}};
	\node[stgstate] (011) at (0.6*\incx,1.5*\incy) {\textcolor{black}{011}};
	\node[stgstate] (111) at (1.6*\incx,1.5*\incy) {\textcolor{black}{111}};

	\draw[stgedge] (010) edge (110);
	\end{tikzpicture}
}	\end{minipage}
&\begin{minipage}{0.18\textwidth}\resizebox{\textwidth}{!}{%
\begin{tikzpicture}
	\tikzstyle{stgstate} = [draw,shape=rectangle,font=\small, fill, color=gray!20]
	\tikzstyle{stgedge} = [-latex, thick,font=\sffamily\normalsize\bfseries]

	\def\incx{1.7}
	\def\incy{1.7}
	\colorlet{darkgreen}{green!50!black}

	\node[stgstate] (000) at (0*\incx,0*\incy) {\textcolor{black}{000}};
	\node[stgstate] (100) at (1*\incx,0*\incy) {\textcolor{black}{100}};
	\node[draw,shape=rectangle,font=\small, fill, color=red!60] (001) at (0.6*\incx,0.5*\incy) {\textcolor{black}{001}};
	\node[draw,shape=rectangle,font=\small, fill, color=red!60] (101) at (1.6*\incx,0.5*\incy) {\textcolor{black}{101}};

	\node[stgstate] (010) at (0*\incx,1*\incy) {\textcolor{black}{010}};
	\node[draw,shape=rectangle,font=\small, fill, color=red!60] (110) at (1*\incx,1*\incy) {\textcolor{black}{110}};
	\node[stgstate] (011) at (0.6*\incx,1.5*\incy) {\textcolor{black}{011}};
	\node[stgstate] (111) at (1.6*\incx,1.5*\incy) {\textcolor{black}{111}};

	\draw[stgedge] (000) edge (010);
	\draw[stgedge] (100) edge (110);
	\draw[stgedge] (011) edge (001);
	\draw[stgedge] (111) edge (101);
\end{tikzpicture}
} \end{minipage}
&\begin{minipage}{0.18\textwidth}\resizebox{\textwidth}{!}{%
\begin{tikzpicture}
	\tikzstyle{stgstate} = [draw,shape=rectangle,font=\small, fill, color=gray!20]
	\tikzstyle{stgedge} = [-latex, thick,font=\sffamily\normalsize\bfseries]

	\def\incx{1.7}
	\def\incy{1.7}
	\colorlet{darkgreen}{green!50!black}

	\node[stgstate] (000) at (0*\incx,0*\incy) {\textcolor{black}{000}};
	\node[stgstate] (100) at (1*\incx,0*\incy) {\textcolor{black}{100}};
	\node[draw,shape=rectangle,font=\small, fill, color=red!60] (001) at (0.6*\incx,0.5*\incy) {\textcolor{black}{001}};
	\node[draw,shape=rectangle,font=\small, fill, color=red!60] (101) at (1.6*\incx,0.5*\incy) {\textcolor{black}{101}};

	\node[stgstate] (010) at (0*\incx,1*\incy) {\textcolor{black}{010}};
	\node[draw,shape=rectangle,font=\small, fill, color=red!60] (110) at (1*\incx,1*\incy) {\textcolor{black}{110}};
	\node[stgstate] (011) at (0.6*\incx,1.5*\incy) {\textcolor{black}{011}};
	\node[stgstate] (111) at (1.6*\incx,1.5*\incy) {\textcolor{black}{111}};

	
	\draw[stgedge] (000) edge (001);
	\draw[stgedge] (100) edge (101);
	\draw[stgedge] (011) edge (010);
	\draw[stgedge] (111) edge (110);
\end{tikzpicture}
} \end{minipage}\\
&&\\\hline
\end{tabular}
\caption{\label{fig:indiv_STG} The 3 independent transition graphs for the Boolean model presented in Figure \ref{fig:toyBN}: $\EE_i=(\BB^3, \mathcal{T}_{i}),\, i=1,2,3$.}
\end{figure}
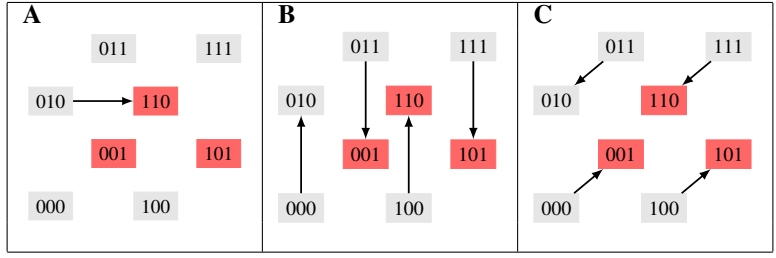

As mentioned in Section \ref{sec:BN}, the complete parametrization of a Boolean Network (BN) $\RR=(G,R,\FF)$, with $|G|=n$, involves selecting a regulatory (Boolean) function for each component in $G$. When considering an asynchronous updating, the STG $\EE = (\BB^n, \mathcal{T})$ representing the complete dynamics of the BN results from the superposition of $n$ independent STGs $\{\EE_i = (\BB^n, \mathcal{T}_i)\}_{i=1,\ldots,n}$ defined on the same set of states $\BB^n$, but where each graph $\EE_i$ encompasses the sole transitions affecting the component $g_i$ as defined by its regulatory function (see Figure \ref{fig:indiv_STG}). 

It is noteworthy that, while the total number of model components may be large, the cardinal of $G_i$, {\it i.e.} the set of regulators of $g_i$, is generally limited (rarely greater than 5). 
Moreover, the regulatory function $f_{i}$ of the component $g_i$ has exactly $|G_i|$ essential variables (conveying the values of the regulators of $g_i$), and consequently $\EE_{i} = (\BB^n, \mathcal{T}_{i})$ can be completely characterized from another STG defined in a reduced state space $\BB^{|G_i|+1}$.

For example, considering components $g_2$ or $g_3$, in our model of Figure \ref{fig:toyBN}, one could work in a $2$-dimensional state space and then project it on the whole $3$-dimensional state space to obtain $\EE_{2}$ or $\EE_{3}$ as displayed in Figure \ref{fig:indiv_STG}. 


\subsection{\label{subsec:consist_f} Characterizing consistent regulatory functions}

Let $\RR=(G,R,\FF)$ be a BN and let us consider $g_i$ with  $G_i$ its set of $p$ regulators. 
Without loss of generality, the regulators of $g_i$ are assumed to be the first $p$ components of $G$: $G_i=\{g_1, \ldots, g_p\}$. 

There are $2^{2^p}$ Boolean functions over the $p$ variables associated to the regulators of $g_i$. However, this huge number can be restricted to some extent, by retaining only the regulatory functions that comply with the regulatory structure of $g_i$, {\it i.e.} that reflect the signs and functionalities of the regulations affecting $g_i$ \cite{abou-jaoude2016}. 

We recall that interaction $(g_j,g_i)$ is functional and positive ({\it i.e.} $g_j\in G_i$ is functional and is an activator) iff:
$$\exists \bold{s} \in \BB^n \mbox{ such that } s_j=0 \mbox{ and }f_i(\bold{s})=\neg f_i(\overline{\bold{s}}^j)=1,$$
where $\overline{\bold{s}}^j$ denotes the state that differs from $\bold{s}$ only in its $j^{th}$ component: $\forall k=1,\ldots, p, k\neq j, s_k=\overline{s}^j_k$ and $s_j=\neg\overline{s}^j_j$. 
Similarly, $(g_j,g_i)$ is functional and negative ({\it i.e.} $g_j$ is functional and is an inhibitor) iff:
$$\exists \mathbf{s} \in \BB^n \mbox{ such that }s_j=1 \mbox{ and }f_i(\bold{s})=\neg f_i(\overline{\bold{s}}^j)=0.$$ 

In other words, if $s_j$ is an essential variable of $f_i$, the interaction $(g_j,g_j)$ is functional; its sign then depends on the values of $f_i$ when switching the value of $s_j$.

Note that we consider the restricted class of BN for which there are no dual regulations, {\it i.e.}, all the regulators are either activators or inhibitors: 
$$
\begin{array}{ll}
\nexists (\bold{s},\bold{s}^\prime) \in \BB^{2n} \mbox{ such that }
&\left\{\begin{array}{l}
s_j=s_j^\prime,\\
f_i(\bold{s})=\neg f_i(\overline{\bold{s}}^j)=1,\\
 f_i(\bold{s}^\prime)=\neg f_i(\overline{\bold{s}^\prime}^j)=0.
 \end{array}\right.
 \end{array}$$

The set of regulators $G_i$ can thus be partitioned as $G_i = G^{+}_i \cup G^{-}_i$, where $G^{+}_i$ is the set of positive regulators of $g_i$ (activators), while components in $G^{-}_i$ are negative regulators of $g_i$ (inhibitors). 

Considering the example in Figure \ref{fig:toyBN}, we have the following sets of regulators: $G_{1} = \{g_1,g_2,g_3\}$, $G^{+}_{1} = \{g_1,g_2\}$ and $G^{-}_{1} = \{g_3\}$, $G_{2} = G^{-}_{2} = \{g_3\}$, $G_{3} = G^{-}_{3} = \{g_2\}$ and $G^{+}_{2} = G^{+}_{3} = \emptyset$. 

Given the component $g_i$, let $\mathcal{F}_i$ be the set of all the {\it consistent Boolean regulatory functions}, {\it i.e.} the functions that comply with the regulatory structure defined by ($G^{+}_i,G^{-}_i$). The following proposition characterizes $\mathcal{F}_i$.

\begin{prop}\label{prop:consistency}

The set $\mathcal{F}_i$ of consistent Boolean regulatory functions of component $g_i$ is the set of nondegenerate monotone Boolean functions $f_i$ such that, $f_i$ is positive in $s_k$ for $g_k \in G^{+}_i$ and negative in $s_k$ for $g_k \in G^{-}_i$.

\end{prop}

Monotonicity derives from the non-duality assumption (an interaction is either positive or negative), and the sign of the interaction from a regulator $g_k$ enforces the positiveness (if $g_k \in G^{+}_i$) or negativeness (if $g_k \in G^{-}_i$). Finally, regulatory functions must be nondegenerate due to the requirement of the functionality of all $g_k\in G_i$.

For the remainder of the paper, we assume that functions in $\mathcal{F}_i$ are represented in a Disjunctive Normal Form ({DNF}):
\begin{equation} \label{eq:DNF}
\forall f_i\in \mathcal{F}_i,\, f_i = C_1 \vee \ldots \vee C_m,
\end{equation}
with,
\begin{equation} \label{eq:conj_clause}
C_j = \bigwedge_{k \in E_j} u_k \qquad j = \{1,\ldots, m\},
\end{equation}
where $E_j \subseteq \{1, \ldots ,p\}$ is the set of indices $k$ such that $s_k$ appears in $C_j$ (recall that $p=|G_i|$). 

The Complete {DNF} ({CDNF}) representation of a consistent Boolean function $f_i$ satisfies the following conditions:
\begin{itemize}
\item[] (i) $\forall g_k \in G_i, \exists j \text{ for which } k \in E_j$;
\item[] (ii) 
$\forall C_j,\,\forall k\in E_j,\,u_k =
\left\{
  \begin{array}{ll}
    s_k , & \hbox{if $g_k \in G^{+}_i$,} \\
    \neg s_k , & \hbox{if $g_k \in G^{-}_s$.}
  \end{array}
\right.$
\end{itemize} 

Both conditions $(i)$ and $(ii)$ derive directly from Proposition \ref{prop:consistency}: $(i)$ stems from the functionality of all  regulators in $G_i$; $(ii)$ guarantees the consistency of the function with the sign of the regulatory interaction $(g_k,g_i)$ (recall that there are no dual regulations); a third condition, which is implicit from the CDNF representation, is that there are no $E_j$, $E_l$ ($j \neq l$) such that $E_j \subset E_l$.

For the BN of Figure \ref{fig:toyBN}, the function $f_{1}(\mathbf{s}) = s_1 \vee (s_2 \land \neg s_3)$ is an element of $\mathcal{F}_{1}$.

Given a ({\em consistent}) regulatory function $f_i \in F_i$, its unique {CDNF} representation can be trivially computed from any DNF representation of $f_i$ by appropriately erasing literals \cite{blake37,crama_hammer}. 

Let $C_j$ be a clause in the {CDNF} representation of $f_i$. Then, the set of states satisfying $C_j$ (true states of $f_i$) can be associated to a subspace of $\BB_i$, as in \cite{Klarner2015}, where $s_k$ is a fixed (resp. free) variable iff $k \in E_j$ (resp. $k \notin E_j$). We call dimension of the subspace associated to a clause $C_j$ (as defined in Eq. \ref{eq:conj_clause}), the number $p - |E_j|$ of free variables of $C_j$. The set $\TT(f)$ of true states of $f$ can then be seen as the union of $m$ subspaces of $\BB^p$ with dimensions $p - |E_j|$, $j = 1,\ldots,m$. 

Given the regulatory structure defined by $G_i$, any function $f_i \in \mathcal{F}_i$ is unambiguously represented by its {set-representation}, as defined below.

\begin{defin}

Given a component $g_i$ with $G_i = G^{+}_i \cup G^{-}_i$ its set of $p$ regulators, the set-representation $S(f_i) \subseteq 2^{\{1, \ldots, p\}}$ of a regulatory function $f_i \in \mathcal{F}_i$ is such that $E_j \in S(f_i)$ if and only if $C_j$ is a conjunctive clause of the CDNF representation of $f_i$ (following notation of Eq. \ref{eq:conj_clause}). 

\end{defin}

In the definition above, $S(f_i)$ represents the structure of $f_i$ as its elements indicate which variables (regulators) are involved in each of the clauses defining $f_i$. The literals (non-complemented and complemented variables) are then unambiguously determined by $G^{+}_i$ and $G^{-}_i$. For example, the set-representation of $f_{1} = s_1 \vee (s_2 \land \neg s_3)$ is $S(f_1) = \{\{1\},\{2,3\}\}$.

Since elements of $S(f_i)$ are pairwise incomparable subsets of $\{1,\ldots, p\}$, for the $\subseteq$ relation, it is easy to verify that $S(f_i)$ is an antichain in the {PO-Set} $(2^{\{1,\ldots, p\}},\subseteq)$. Moreover, $S(f_i)$ is also a cover of $\{1,\ldots, p\}$ since all indices in $\{1,\ldots, p\}$ have to be present in at least one element of $S(f_i)$. Finally, any antichain in $(2^{\{1,\ldots, p\}},\subseteq)$ which is a cover of $\{1,\ldots, p\}$ {is the set representation} of a unique function in $\mathcal{F}_i$. Therefore, $\mathcal{F}_i$ is isomorphic to the set $\mathcal{S}_p$ of antichains in $(2^{\{1,\ldots, p\}},\subseteq)$. 

\begin{figure*}[tbh]
\begin{tabular}{lr}
\begin{minipage}{0.48\textwidth}\resizebox{\textwidth}{!}{
\begin{tabular}{c|r|r}
		$p$ & \multicolumn{1}{c}{$M(p)$} &\multicolumn{1}{c}{$N(p)=|\mathcal{F}_g| = |\mathcal{S}_p|$}\\ \hline
		1 & 3&1\\
		2 &6& 2\\
		3 & 20&9\\
		4 & 168&114\\
		5 & 7 581&6 894\\
		6 & 7 828 354&7 785 062\\
		7 & 2 414 682 040 998 & 2 414 627 396 434\\
		8& 56 130 437 228 687 557 907 788& 56 130 437 209 370 320 359 968
	\end{tabular}}
\end{minipage}
&\begin{minipage}{0.48\textwidth}
\includegraphics[width=\textwidth]{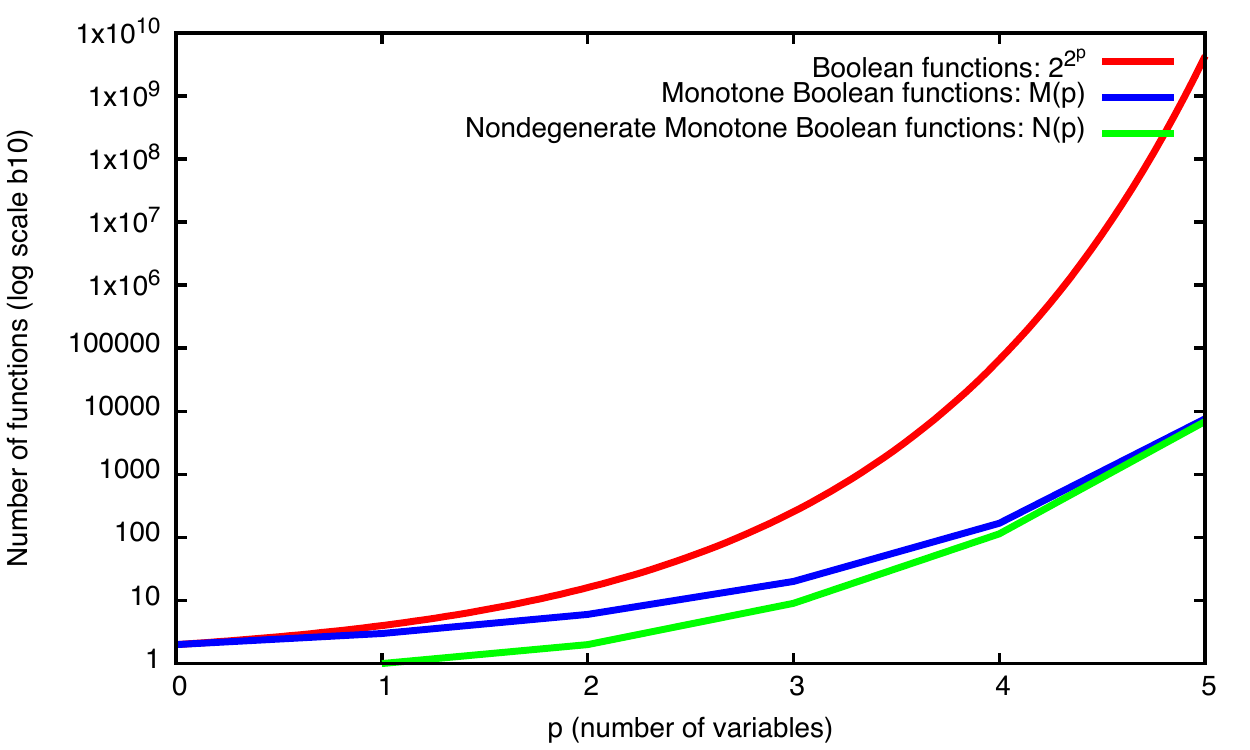}
\end{minipage}
\end{tabular}
	\caption{Left: number of monotone Boolean functions (Dedekind number $M(p)$) and of nondegenerate monotone Boolean functions ($N(p)$) of $p=1,\ldots, 8$ variables. $N(p)$ is also the number of antichain covers of $\{1,\ldots, p\}$. Right: the plots illustrate the growth pattern of these numbers, including that of the total number of Boolean functions\label{fig:nfunctions}.}
\end{figure*}

The cardinality $N(p)$ of $\mathcal{F}_i$, set of all nondegenerate monotone Boolean functions of $p$ variables, is smaller than $2^{2^{p}}$, the number of all Boolean functions of $p$ variables and also than $M(p)$, the Dedekind number of monotone Boolean functions (including degenerate functions). Indeed, one can easily show that:
 $$N(p)=M(p)-2-\sum_{k=1}^{p-1}\frac{p!}{k!(p-k)!}N(k).$$

Nevertheless, as illustrated in Figure \ref{fig:nfunctions}, the cardinality of $\mathcal{F}_i$ dramatically increases with the number of variables (regulators of $g_i$) and thus constitutes a major computational challenge.
Any approach relying on the exploration of the full set $\mathcal{F}_i$ where $g_i$ has more than 5 regulators, would be intractable. In this context, the characterization of the structure of $\mathcal{F}_i$ might be helpful to assess the impact of particular regulatory functions on the dynamics of the corresponding BN.

\subsection{Structuring $\mathcal{F}_i$ as a Partially Ordered Set}

In this section, we show that given a component $g_i$, the set of its consistent regulatory functions $\mathcal{F}_i$   can be structured as a {PO-Set}. To this end, let consider the binary relation $\preceq$ on $\mathcal{F}_i \times \mathcal{F}_i$ defined by: 

$$\forall f,f^\prime \in \mathcal{F}_i,\, f \preceq\ f^{\prime} \iff \TT(f) \subseteq \TT(f^{\prime}).$$

\begin{figure}[!tpb]
	\centering
	\resizebox{0.5\textwidth}{!}{%
\begin{tikzpicture}
  \tikzstyle{hdstate} = [ellipse,draw,blue,font=\bfseries\small]
	\tikzstyle{hdedge} = [gray,font=\sffamily\normalsize\bfseries]
	\tikzstyle{hdfunc} = [red,font=\large]

	\def\xspace{3.5}
	\def\yspace{2}
	
	\node[hdstate] (1-2-3) at (0,4*\yspace) {\{\{1\},\{2\},\{3\}\}};

	\node[hdstate] (3-12) at (-1*\xspace,3*\yspace) {\{\{3\},\{1,2\}\}};
	\node[hdstate] (2-13) at (0,3*\yspace) {\{\{2\},\{1,3\}\}};
	\node[hdstate] (1-23) at (1*\xspace,3*\yspace) {\{\{1\},\{2,3\}\}};

	\node[hdstate] (12-13-23) at (0,2*\yspace) {\{\{1,2\},\{1,3\},\{2,3\}\}};

	\node[hdstate] (12-23) at (-1*\xspace,1*\yspace) {\{\{1,2\},\{2,3\}\}};
	\node[hdstate] (12-13) at (0,1*\yspace) {\{\{1,2\},\{1,3\}\}};
	\node[hdstate] (23-13) at (1*\xspace,1*\yspace) {\{\{1,3\},\{2,3\}\}};

	\node[hdstate] (123) at (0,0) {\{\{1,2,3\}\}};

	\draw[hdedge] (3-12) edge node[black] {R3} (1-2-3);
	\draw[hdedge] (2-13) edge node[black] {R3} (1-2-3);
	\draw[hdedge] (1-23) edge node[black] {R3} (1-2-3);

	\draw[hdedge] (12-13-23) edge node[black] {R2} (3-12);
	\draw[hdedge] (12-13-23) edge node[black] {R2} (2-13);
	\draw[hdedge] (12-13-23) edge node[black] {R2} (1-23);

	\draw[hdedge] (12-23) edge node[black] {R1} (12-13-23);
	\draw[hdedge] (12-13) edge node[black] {R1} (12-13-23);
	\draw[hdedge] (23-13) edge node[black] {R1} (12-13-23);

	\draw[hdedge] (123) edge node[black] {R3} (12-23);
	\draw[hdedge] (123) edge node[black] {R3} (12-13);
	\draw[hdedge] (123) edge node[black] {R3} (23-13);

	\path[hdfunc] (1-2-3) ++(1.5,0.6) node {sup $F_{g_1} = s_1 \vee s_2 \vee \neg s_3$};
	\path[hdfunc] (123) ++(1.5,-0.6) node {inf $F_{g_1} = s_1 \wedge s_2 \wedge \neg s_3$};
	\path[hdfunc] (1-23) ++(1,0.6) node {$f^{1}_{g_1} = s_1 \vee (s_2 \wedge \neg s_3)$};
\end{tikzpicture}
	}%
	\caption{Hasse Diagram representing the set of all possible functions composed of 3 regulators ({\it e.g.} functions in red of the component $g_1$ of the model in Figure \ref{fig:toyBN}). R1, R2 and R3 labels indicate the corresponding rule applied to compute the neighboring parent/child node (see Section \ref{sec:neighbors}).}\label{fig:HD_3}
\end{figure}
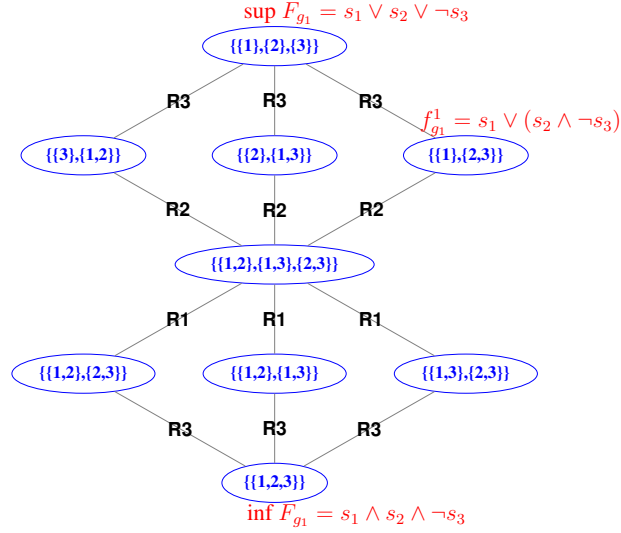

It is easy to verify that $(\mathcal{F}_{i},\preceq)$ is a {PO-Set}. Figure \ref{fig:HD_3} shows the Hasse Diagram (HD) of the {PO-Set} $(\mathcal{F}_{1},\preceq)$ of  $g_1$, component of the model presented in Figure \ref{fig:toyBN}. This {PO-Set} has Supremum and Infimum elements given by $\emph{sup }\mathcal{F}_{1} = s_1 \vee s_2 \vee \neg s_3$ and ${inf }\mathcal{F}_{1} = (s_1 \land s_2 \land \neg s_3)$, respectively, with  $S_{sup}=\{\{1\},\{2\},\{3\}\}$ and $S_{inf}=\{\{1,2,3\}\}$ as set-representations.

Observe that, while the functions in $\mathcal{F}_i$ depend on the specific regulatory structure ({\it i.e.}, the signs of the regulations), the topology of the {HD} and the relation between its nodes, when seen as set-representations, only depend on $p$, the number of regulators of $g_i$.  In other words, the HD shown in  Figure \ref{fig:HD_3} represents the set of consistent regulatory functions for any component with 3 regulators.

In fact, one can consider the relation $\preceq$ on the set $\mathcal{S}_p$ of antichains in $(2^{\{1,\ldots, p\}},\subseteq)$:
$$\forall S, S^\prime \in \mathcal{S}_p,\, S\preceq S^{\prime} \iff \forall \sigma \in S,\, \exists \sigma^{\prime} \in S^{\prime} \text{ such that } \sigma^{\prime} \subseteq \sigma.$$

Recall that $(\mathcal{S}_p,\preceq)$ is also a {PO-Set}. Its {HD} has the same strucure of  the HD of $(\mathcal{F}_{g},\preceq)$, where its nodes are the set-representations $S(f) \in \mathcal{S}$. This is a because:
\begin{equation} \label{eq:two_relations}
f \preceq f^{\prime} \iff S(f) \preceq S(f^{\prime}).
\end{equation}

Furthermore, the set-representation of a function $f$ in $\mathcal{F}_i$ is sufficient to determine the number and \emph{signatures} of its true states (elements $\TT(f)$), independently of the signs of the $p$ interactions targeting $g_i$. We introduce the notion of {signature} of a state $\bold{s} \in \BB^n$ as a $p$-tuple $v$ composed of symbols in $\{o,\overline{o},\star\}$ such that, $\forall k=1,\ldots, p$: $v_k = o$ means that $s_k = 1$ if $g_k\in G^{+}_i$ and $s_k = 0$ if $g_k \in G^{-}_i$ ({\it i.e.}, the regulation from $g_k$ is operative); $v_k = \overline{o}$ otherwise; the symbol $\star$ means that both values ($0$ and $1$) are admissible. 

Given a function $f$ in $\mathcal{F}_i$ and its set-representation $S(f)$, the signatures of the true states of $f$ are obtained from the subsets of $S(f)$ as follows: given $\sigma \in S(f)$, if $k \in \sigma$ then $v_k = o$ if $g_k\in G^+_i$, $v_k = \overline{o}$ if $g_k\in G^-_i$, and otherwise $v_k= \star$, which accounts for both $o$ and $\overline{o}$.


For example, in the {HD} of Figure \ref{fig:HD_3}, consider the set $S({\inf }(\mathcal{F}_1))=\{\{1,2,3\}\}$ and the regulatory structure given by $G_1^+=\{g_1,g_2\}$ and $G_1^-=\{g_3\}$. These altogether define the signature $(o,o,\overline{o})$ of the elements of $\TT(\inf (\mathcal{F}_1))$ that in turn defines the sole state $\bold{s}=(1,1,0)$.
For the function $f=(s_1 \land s_2) \vee (s_1 \land \neg s_3)$, its set-representation is $\{\{1,2\},\{1,3\}\}$ defines the signatures $(o,o,\star)\text{ and }(o,\star,\overline{o})$ (or $(o,o,o),(o,o,\overline{o})\text{ and }(o,\overline{o},\overline{o})$), which in turn specify the set of true states $\TT(f)=\{(1,1,0),(1,1,1),(1,0,0)\}$.

Summarising, the {PO-Set} $(\mathcal{S}_p,\preceq)$ can be used as a template for all {PO-Sets} $(\mathcal{F}_i,\preceq)$ of regulatory functions of a component $g_i$ with $p$ regulators, considering any possible regulatory structures, {\it i.e.}, all pairs $(G^{+}_i,G^{-}_i)\in G$. In what follows, properties of {PO-Sets} $(\mathcal{F}_{i},\preceq)$ will thus be derived from those of $(\mathcal{S}_p,\preceq)$.

It is easy to verify that the {PO-Sets} $(\mathcal{F}_{g},\preceq)$ and $(\mathcal{S}_p,\preceq)$ are bounded {PO-Sets}, their Supremum being the regulatory function \emph{presence of at least one activator or absence of at least one inhibitor}, and their Infimum being the function \emph{presence of all activators and absence of all inhibitors}. On the other hand, although when $|G_g| \leq 3$ both {PO-Sets} are clearly \emph{Complete Lattices}, in general this is not true for larger number of regulators. For example, for $p = 4$, if one considers $S_1 = \{\{3\},\{1,2,4\}\}$ and $S_2 = \{\{2,3\},\{1,4\}\}$, then $S = \{\{1,2\},\{3\},\{4\}\}$ and $S^{\prime} = \{\{3\},\{2,4\},\{1,4\}\}$ are both minimal upper bounds of $\{S_{1},S_{2}\} \subset\mathcal{S}$ so that $\{S_{1},S_{2}\}$ has not a least upper bound in $(\mathcal{S}_4,\preceq)$.


\section{Characterizing the vicinity of elements of the PO-Set$(\mathcal{F}_{i},\preceq)$}\label{sec:neighbors}


Given a generic component $g_i$ with $p$ regulators, we first introduce some terminology on the relationships between elements in the HD of the {PO-Set} $(\mathcal{S}_p,\preceq)$ (obviously, this terminology also applies to $(\mathcal{F}_i,\preceq)$). Given $S, S^\prime \in \mathcal{S}_p$:
\begin{itemize}
\item $S^\prime$ is a \emph{parent} of $S$ in $(\mathcal{S}_p,\preceq)$ if and only if $S \preceq S^\prime$ and $\nexists S^{\prime\prime} \in \mathcal{S}$ such that $S \preceq S^{\prime\prime}$ and $S^{\prime\prime} \preceq S^\prime$;
\item $S^\prime$ is a \emph{child} of $S$ if and only if $S$ is a {parent} of $S^\prime$;
\item $S^\prime$ is a \emph{sibling} of $S$ if and only if it shares a common parent with $S$;
\item $S^\prime$ is a \emph{direct neighbor} of $S$ if and only if it is either a {parent} or a {child} of $S$.
\end{itemize}

For example, the set $\{\{1\},\{2,3\}\}$ in $\mathcal{S}_3$ has a unique parent $\{\{1\},\{2\},\{3\}\}$, a unique child $\{\{1,2\},\{1,3\},\{2,3\}\}$ (the two are the direct neighbors of the set), and two siblings $\{\{2\},\{1,3\}\}$ and $\{\{3\},\{1,2\}\}$ (see Figure \ref{fig:HD_3}).


The following two sets of rules allow us to compute, for any element of the {PO-Set} $(\mathcal{S}_p,\preceq)$, the set of its direct neighbors ({parents} and {children}).

\begin{rulesparents}
Given an element $S$ of the {PO-Set} $(\mathcal{S}_p,\preceq)$, a parent $S^\prime$ of $S$ is obtained by applying one of the following rules:

\begin{enumerate}
\item $S^\prime = S \cup \{c\}$, with element $c \in \max(\{\sigma \subseteq \{1,\ldots p\}$ such that $\sigma \text{ is \emph{independent} of } S\})$; 

\item $S^\prime = \min(S \cup \{\sigma\})$ with $\sigma\subseteq \{1,\ldots,p\}$ such that: 
\begin{enumerate}
\item $\exists \sigma^\prime \in S$ such that $\sigma \subset \sigma^\prime$;
\item $\nexists \sigma^{\prime} \subseteq \{1,\ldots,p\}$ and $\sigma^{\prime\prime} \in S$ such that $\sigma \subsetneq\sigma^{\prime} \subsetneq\sigma^{\prime\prime}$;
\item $\sigma \nsubseteq c$, $\forall c$ satisfying rule 1;
\item $S^\prime$ is a cover of $\{1,\ldots,p\}$;
\end{enumerate}

\item $S^\prime = \min(S \cup \{\sigma\} \cup \{\sigma^{\prime}\})$, with $\sigma,\,\sigma^\prime$ subsets of $\{1,\ldots, p\}$ such that:

\begin{enumerate}
\item $\sigma$ and $\sigma^{\prime}$ satisfy all the conditions of rule 2 but condition (c);
\item $S^\prime$ is a cover of $\{1,\ldots,p\}$.
\end{enumerate}
\end{enumerate}
\end{rulesparents}

\begin{ruleschildren}
Given an element $S$ of the {PO-Set} $(\mathcal{S}_p,\preceq)$, a child $S^\prime$ of $S$ is obtained by applying one of the following rules:
\begin{enumerate}
\item $S^\prime = S \setminus \{c\}$ with $c$ such that $\nexists \sigma$ independent of $S \setminus \{c\}$ such that $c \subset \sigma$;
\item $S^\prime = (S \setminus \{c\}) \cup C$, for any $c$ and $C$ such that:
\begin{enumerate}
\item $C = \min(\{ \sigma\;|\;\sigma\text{ independent of }(S \setminus \{c\})\text{ and } \sigma \supset c\})$
\end{enumerate}
\item $S^\prime = (S \setminus \{c,c^\prime\}) \cup \{c \cup c^\prime\}$ for $c,c^\prime$ not satisfying rules 1 or 2 and such that $c \cap c^\prime \neq \emptyset$.
\end{enumerate}
\end{ruleschildren}
\medskip

\begin{theorem}

Given an element $S$ of the {PO-Set} $(\mathcal{S}_p,\preceq)$, $S^\prime$ is a parent (resp. a child) of $S$ if and only if $S^\prime$ is generated by a \emph{Rule to compute parents} (resp. a \emph{Rule to compute children}).

\end{theorem}

\begin{proof}
We start by considering the \emph{Rules to compute parents}.

Let $S = \{c_1, \ldots, c_m\}$ be any element of the {PO-Set} $(\mathcal{S}_p,\preceq)$. 
Let us write $S^\prime = \{c^{\prime}_1, \ldots, c^{\prime}_{m^\prime}\}$ for a parent of $S$. 
First, by definition, $S^\prime$ is a parent of $S$ if and only if: (a) $S^\prime \ne S$, (b) for each $c_i$, $i \in \{1,\ldots,m\}$, there exists at least one $c^{\prime}_{j}$, $j \in \{1,\ldots,m^{\prime}\}$, with $c^{\prime}_{j} \subseteq c_i$, and (c) there is no $S^{\prime\prime}$ such that $S \preceq S^{\prime\prime} \preceq S^\prime$.

\begin{itemize}
\item[(i)] First, let us consider the case where there is no pair $i,j$ such that $c^{\prime}_{j}$ is a proper subset of $c_i$. Then, it is clear that for all $i \in \{1,\ldots,m\}$ there is a $j \in \{1,\ldots,m^\prime\}$ such that $c_i = c^{\prime}_{j}$ and consequently, $m < m^\prime$ and $S^\prime$ can be written, without loss of generality, $S^\prime = S \cup \{c^{\prime}_{m+1},\ldots,c^{\prime}_{m^\prime}\}$. Now, it is easy to see that the set $\{c^{\prime}_{m+1},\ldots,c^{\prime}_{m^\prime}\}$ must be a singleton, otherwise we could build $S^{\prime\prime}$ with $S \preceq S^{\prime\prime} \preceq S^\prime$ by setting $S^{\prime\prime} = S \cup \{c^{\prime}_{j}\}$, with $j\in\{m+1,\ldots,m^\prime\}$. Thus $S^\prime$ can be written $S^\prime = S \cup \{c\}$. 

We now show that in this case, the necessary and sufficient conditions for $S^\prime$ to be a parent of $S$ are those stated in \emph{Rule-1 to compute parents}. First, it is clear that $c$ has to be independent of $S$, otherwise $S^\prime$ would not be an element of $(\mathcal{S}_p,\preceq)$. Second, there cannot exist a $\sigma^\prime$ independent of $S$ such that $\sigma^\prime \supset c$, otherwise $S^{\prime\prime} = S \cup \{\sigma^\prime\}$ would be such that $S \preceq S^{\prime\prime} \preceq S^\prime$. Hence the condition for $S^\prime$ to be a parent of $S$ is that $c \in Max(\{\sigma^\prime \subseteq N \;|\; \sigma^\prime \text{ is \emph{independent} of } S\})$.

\item[(ii)] Let us now consider the case where there is at least one pair $i,j$ where $c^{\prime}_{j}$ is a proper subset of $c_i$. Then statements of \emph{Rule 2-(b,c,d)} are necessary and sufficient conditions for $S^\prime$ to be a parent of $S$. In fact, \emph{Rule 2-(b,c)} are conditions for not having $S^{\prime\prime}$ with $S \preceq S^{\prime\prime} \preceq S^\prime$. On the other hand, \emph{Rule 2-(d)} is the condition for $S^\prime$ to be in $(\mathcal{S}_p,\preceq)$.

\item[(iii)] Finally, case 3 corresponds to the situation where an additional subset has to be added such that $S^\prime$ is a cover of $\{1,\ldots,p\}$.
\end{itemize}
The proof of the \emph{Rules to compute children} case follows from the observation that those rules are counterparts of the \emph{Rules to compute parents} by simply changing the roles of $S$ and $S^\prime$.

\end{proof}


The following proposition concerns the difference in the number of true states for two direct neighbors in $(\mathcal{F}_g,\preceq)$. Recall that $|G_g|=p$ ($g$ has $p$ regulators).

\begin{prop}\label{prop:st_between}

Let $S,S^\prime \in \mathcal{S}$ be such that $S^\prime$ is a parent of $S$ in $(\mathcal{S}_p,\preceq)$, and let $f,f^\prime \in \mathcal{F}_g$ be the corresponding functions in $(\mathcal{F}_{g},\preceq)$. Thus, $\TT(f) \subset \TT(f^\prime)$ and $|\TT(f^\prime) \setminus \TT(f)| \in \{1,2\}$.

\end{prop}

The proof of this proposition uses the following auxiliary result.

\begin{lem}\label{lem:st_bet_cl}

Let $\sigma \subseteq \{1,\ldots,p\}$ and $\alpha = Min(\{\beta \subseteq \{1,\ldots,p\} \;|\; \beta \supset \sigma\})$. Let $f_\sigma$ and $f_{\alpha}$ be monotone Boolean functions having respectively $\sigma$ and $\alpha$ as set-representations. There is only one state $\mathbf{s} \in \TT(f_{\sigma}) \setminus \TT(f_{\alpha})$, namely, the one with signature $v = (v_{1}, \dots,v_{p})$, where $v_{k} = o$ for $k \in \sigma$ and $v_{k} = \overline{o}$ for $k \notin \sigma$.

\end{lem}

\begin{proof}

The signature of the set of states $\TT(f_{\sigma})$ is $V_\sigma = \{v \in \{o,\overline{o}\}^{p} \;|\; v_k = o \text{ for } k \in \sigma \;and\; v_k = * \text{ for } k \notin \sigma\}$. On the other hand, observe that $\alpha$ is the set of ($p - |\sigma|$) different subsets $\beta \subseteq \{1,\ldots,p\}$ with $\beta \supset \sigma$ and $|\beta| = |\sigma| + 1$. Each of those subsets represents a clause of the DNF of $f_{\beta}$. The signature of the set of states $\TT(f_{\beta})$ is $V_{\alpha} = \{v \in \{o,\overline{o}\}^{p} \;|\; v_k = o \text{ for } k \in \sigma \text{ and for one } k \notin \sigma\}$. Therefore the signature of $\TT(f_{\sigma}) \setminus \TT(f_{\alpha})$ is $V_\sigma \setminus V_{\alpha} = \{v \in \{o,\overline{o}\}^{p} \;|\; v_k = o \text{ for } k \in \sigma \;and\; v_k = \overline o \text{ for all } k \notin \sigma\}$.

\end{proof} 

\begin{proof}[Proof of Proposition \ref{prop:st_between}]

We will consider the three different ways of generating a parent $S^\prime$ for $S$.\\
\begin{itemize}
\item Using \emph{Rule 1}, it is clear that the set of states in $\TT(f^\prime) \setminus \TT(f)$ is composed by the states verifying the clause represented by the subset $c$ and not in $\TT(f)$. Observe that this set is necessarily non-empty, otherwise $c$ would not be independent of $S$. Since $c \in \emph{Max}(\{\sigma^\prime \subseteq \{1,\ldots,p\} \;|\; \sigma^\prime \text{ is independent of } S\})$ it follows that $Min(\{\beta \subseteq \{1,\ldots,p\} \;|\; \beta \supset c\})$ are all dependent of $S$, which means that all the states satisfying the clauses represented by this set are already in $\TT(f)$. Thus, by lemma \ref{lem:st_bet_cl}, there is only one state in $\TT(f^\prime) \setminus \TT(f)$.

\item Using \emph{Rule 2}, $\sigma^{\prime}$ replaces all subsets $Min(\{\beta \subseteq \{1,\ldots,p\} \;|\; \beta \supset \sigma^{\prime}\})$ in $S$. Thus, again by lemma \ref{lem:st_bet_cl}, there is only one state in $\TT(f^\prime) \setminus \TT(f)$.

\item Using \emph{Rule 3}, the same reasoning shows that each $\sigma^{\prime}$ and $\sigma^{\prime\prime}$ introduce one state in $\TT(f^\prime) \setminus \TT(f)$, leading in this case to $|\TT(f) \setminus \TT(f^\prime)| = 2$.
\end{itemize}
\end{proof}

Summarizing, $|\TT(f^\prime) \setminus \TT(f)| = 1$ when \emph{Rules 1 or 2 to compute parents} apply to relate $S^\prime$ to $S$, and $|\TT(f^\prime) \setminus \TT(f)| = 2$ when \emph{Rule 3} applies. The same holds when considering the number of true states of a function and that of one of its child.


\section{Assessing how changing the regulatory function impacts the dynamics}\label{sec:impact}
	
\subsection{Number of transitions over component $g$}

In the following, some results are derived concerning the number of transitions in $\EE_{g} = (\BB^p, \mathcal{T}_{f_g})$, depending on the regulatory functions in $\mathcal{F}_{g}$. The total number of transitions and the number of increasing ($\mathcal{T}^{+}_{f_g}$) and decreasing ($\mathcal{T}^{-}_{f_g}$) transitions are considered.

For the sake of simplicity, it will be assumed in all cases that the set of $n$ components $G = \{g_{k}\}_{k = 1, \ldots, n}$ in the Boolean network is equal to the set of $p$ regulators of $g$, plus the component $g$ ($G=G_g\cup\{g\}$). Without loss of generality, it will also be assumed that $g$ is the last component in $G$ ({\it i.e.,} it has the greatest index). If $g$ is auto-regulated ($g\in G_g$), $p = n$ otherwise $n = p + 1$. Extending the results to the case where $p < n$ would be straightforward.


The following Proposition~\ref{prop:transitions} introduces bounds (or invariance) on the total numbers of transitions in $\EE_{g}$, for regulatory functions in $\mathcal{F}_{g}$.

\begin{prop}\label{prop:transitions}

Let $G_g$ be the set of regulators for $g$. For any regulatory function $f_{g} : \BB^n \rightarrow \BB$ in $\mathcal{F}_{g}$, the set of transitions in the resulting STG $\EE_{g} = (\BB^n, \mathcal{T}_g)$ is such that:
\begin{enumerate}
\item $|\mathcal{T}_{f_g}| = 2^{n-1}$ if $g \notin G_g$;
\item $0 \le |\mathcal{T}_{f_g}| < 2^{(n-1)}$ if $g \in G^{+}_g$;
\item $2^{(n-1)} < |\mathcal{T}_{f_g}| \leq 2^{n}$ if $g \in G^{-}_g$.
\end{enumerate}

\end{prop}

\begin{proof}

For $g \notin G_g$ ($g=g_{p+1}$ is not auto-regulated, and $n=p+1$), a state $\mathbf{s} \in \BB^n$ can be written as $\bold{s} = (s_1,\ldots, s_p, s_{p+1})$ and such that any function $f_{g} : \BB^n \rightarrow \BB$ is independent on the value of component $s_{p+1}$ (recall $n=p+1$). Now, for $\mathbf{s}|_{g} = (s_1,\ldots, s_p) \in \BB^p$, if $f_{g}(\mathbf{s}|_{g}) = 0$ then $[(s_1,\ldots, s_p,1), (s_1,\ldots, s_p, 0)] \in \mathcal{T}_{f_g}$ is the only transition with origin in states of the set $\{(s_1,\ldots, s_n, *)\}$. The same reasoning applies to the case where $f_{g}(\mathbf{s}|_{g}) = 1$, and in this case $[(s_1,\ldots, s_n, 0), (s_1,\ldots, s_n, 1)] \in \mathcal{T}_{f_g}$ is the only transition with origin in states of the set. Thus, there is one and only one transition for each pair of states $\{(s_1,\ldots, s_n, 0),(s_1,\ldots, s_n, 1)\}$ irrespectively of its direction. This proves that when $g \notin G_g$, the number of transitions in $\EE_{g}$ is half of the size of the state space, namely, $|\mathcal{T}_{f_g}| = 2^{(p+1)}/2 = 2^{p}$.

Let us now consider the case where $g$ is auto-regulated ($g \in G_g$, and $p=n$). In this case, $f_{g}$ depends also on $s_n$, the state of $g$. There are several possibilities. 

If $g \in G^{+}_g$, for $\mathbf{s} = (s_1,\ldots, s_{n-1}, 0)$, if $f_{g}(\mathbf{s}) = 1$ then $[(s_1,\ldots, s_{n-1}, 0), (s_1,\ldots, s_{n-1}, 1)] \in \mathcal{T}_{f_g}$; besides, in this case for $\mathbf{s^\prime}=(s_1,\ldots, s_{n-1}, 0)$, $f_g(\mathbf{s^\prime})$ cannot take value $0$, because the only component that changes between $\mathbf{s}$ and $\mathbf{s^\prime}$ is $s_n$, and $g$ activates itself; thus $f_{g}(\mathbf{s^\prime}) = 1$ and there is no outgoing transition from $\mathbf{s^\prime} = (s_1,\ldots, s_{n-1}, 1)$. 

Now, still for $g \in G^{+}_g$, for $\mathbf{s} = (s_1,\ldots,s_{n-1}, 0)$, if $f_{g}(\mathbf{s}) = 0$, there is no outgoing transition from $\mathbf{s}$; in this case, for $\mathbf{s^\prime} = (s_1,\ldots, s_{n-1}, 1)$, $f_{g}(\mathbf{s^\prime})$ can take values $0$ or $1$; in the latter case, there is no outgoing transition from $\mathbf{s^\prime}$, while in the former case $[(s_1,\ldots, s_{n-1}, 1), (s_1,\ldots, s_{n-1}, 0)] \in \mathcal{T}_{f_g}$.

This shows that, if $g \in G^{+}_g$, for each pair of states $(s_1,\ldots, s_{n-1}, *)$ there is at most one transition. This would impose an upper bound of $|\BB^n|/2 = 2^{n}/2 = 2^{(n-1)}$ for $|\mathcal{T}_{f_g}|$. Moreover, the only possibility to reach this limit would be if $f_{g}$ did not change its value when $s_n$ changes, for all pairs of states $(s_1, \ldots, s_{n-1}, *)$. But this would imply a non-functional auto-regulation, which contradicts the consistency condition imposed to $\mathcal{F}_{g}$. Thus $|\mathcal{T}_{f_g}| \le 2^{(n-1)}$.

Let us now consider the case for $g \in G^{-}_g$. In this case, $f_{g}$ depends also on $s_n$. For $\mathbf{s} = (s_1,\ldots, s_{n-1}, 0)$, if $f_{g}(\mathbf{s}) = 0$ there is no outgoing transition from $\mathbf{s}$; besides, for $\mathbf{s^\prime} = (s_1,\ldots, s_{n-1}, 1)$, $f_{g}(\mathbf{s^\prime})$ cannot take value $1$ because $g$ represses itself; thus $f_{g}(\mathbf{s^\prime}) = 0$ and $[(s_1,\ldots, s_{n-1}, 1), (s_1,\ldots, s_{n-1}, 0)] \in \mathcal{T}_{f_g}$.

Now, still for $g \in G^{-}_g$, for $\mathbf{s} = (s_1,\ldots, s_{n-1}, 0)$, if $f_{g}(\mathbf{s}) = 1$ then $[(s_1,\ldots, s_{n-1}, 0), (s_1,\ldots, s_{n-1}, 1)] \in \mathcal{T}_{f_g}$; in this case, for $\mathbf{s^\prime} = (s_1,\ldots, s_{n-1}, 1)$, $f_{g}(\mathbf{s^\prime})$ can take values $1$ or $0$; in the latter case $[(s_1\ldots, s_{n-1}, 1), (s_1,\ldots, s_{n-1}, 0)] \in \mathcal{T}_{f_g}$, while in the former case, there is no outgoing transition from $\mathbf{s^\prime} =(s_1,\ldots, s_{n-1}, 1)$.

	This shows that, if $g \in G^{-}_g$, for each pair of states $(s_1,\ldots, s_{n-1}, *)$, there is at least one and at most two transitions. Thus $|\BB^n|/2$ and $|\BB^n|$ are lower and upper bounds for $|\mathcal{T}_{f_g}|$, respectively. But, regarding the lower bound, the only possibility to reach it would be, as previously, if $f_{g}$ did not change its value with $s_n$, for all pairs $(s_1,\ldots, s_{n-1}, *)$, and this would lead to the same contradiction in the consistency condition imposed to $\mathcal{F}_{g}$.

In summary, the number of transitions in $\EE_{g}$ when $g \in G^{-}_g$ is $|\BB^n|/2 < |\mathcal{T}_{f_g}| \leq |\BB^n|$, that is $2^{(n-1)} < |\mathcal{T}_{f_g}| \leq 2^{n}$.

\end{proof}

Some of the bounds established in Proposition \ref{prop:transitions} deserve further discussion.
In principle, when $g \in G^{+}_g$, the lower bound for $|\mathcal{T}_{f_g}|$ is $0$. There are no transitions in $\EE_{g}$ if and only if $f_{g}(\mathbf{s}) = 1$ for all $\mathbf{s} \in \BB^n$ such that $s_n = 1$ and $f_{g}(\mathbf{s}) = 0$ for all $\mathbf{s}$ such that $s_n = 0$. The only Boolean function satisfying these conditions is $f(\mathbf{s}) = s_n$ for all $\mathbf{s}\in \BB^n$, which is consistent only if $g$ is its sole regulator, {\it i.e.,} $G_g = G^{+}_g = \{g\}$.

Furthermore, a similar reasoning allows to deduce that, when $g \in G^{-}_g$, the upper bound $2^{n}$ of $|\mathcal{T}_{f_g}|$ is reached only if $g$ is its sole regulator, {\it i.e.,} $G_g = G^{-}_g = \{g\}$.

The next proposition establishes bounds for the numbers of increasing and decreasing transitions in $\EE_{g}$, considering regulatory functions in $\mathcal{F}_{g}$.

\begin{prop} \label{prop:incr_decr-trans} 

The upper ($U_g$) and lower ($L_g$) bounds for the numbers of increasing and decreasing transitions in the STG $\EE_{g}$ are:
\begin{enumerate}
\item $U_g = 2^{n-1} - 1$ and $L_g = 1$ if $g \notin G_g$;
\item $U_g = 2^{n - 1} -1$ and $L_g = 0$ if $g \in G^{+}_g$;
\item $U_g = 2^{n - 1}$ and $L_g = 1$ if $g \in G^{-}_g$.
\end{enumerate}

\end{prop}

The following lemma will be used to prove Proposition \ref{prop:incr_decr-trans}.

\begin{lem}\label{lem:true_false}

For any function $f_g : \BB^n \rightarrow \BB$ in the {PO-Set} $(\mathcal{F}_{g},\preceq)$, we have:
\begin{enumerate}
\item $f_g(\mathbf{s})=1$ for $\mathbf{s} \in \BB^n$ with signature $v$ such that $v_k = o$ for all $k \in \{1,\ldots,n\}$;
\item $f_g(\mathbf{s})=0$ for $\mathbf{s} \in \BB^n$ with signature $v$ such that $v_k = \overline{o}$ for all $k \in \{1,\ldots,n\}$.
\end{enumerate}

\end{lem}

\begin{proof}

Let $\emph{inf }\mathcal{F}_{g} = \bigwedge_{k | g_k \in G^{+}_{g}} s_k \bigwedge_{k | g_k \in G^{-}_{g}} \neg s_k$. The only state for which $\emph{inf }\mathcal{F}_{g} = 1$ is $\mathbf{s}$ such that $s_k = 1$ if $g_k \in G^{+}_{g}$ and $s_k = 0$ if $g_k \in G^{-}_{g}$, in other words the state with signature $v$ with $v_k =o$ for all $k \in \{1,\ldots, n\}$. For all $f_g \in \mathcal{F}_{g}$, $\emph{inf }\mathcal{F}_{g} \preceq f_g$ and thus $f_{g}(\mathbf{s}) =1$.

Let us now consider $\emph{sup}\mathcal{F}_{g} = \bigvee_{k | g_k \in G^{+}_{g}} s_k \bigvee_{k | g_k \in G^{-}_{g}} \neg s_k$. This function takes value $1$ for all but one state in $\BB^n$, namely $\mathbf{s}^\prime$ such that $s^\prime_k = 0$ if $g_k \in G^{+}_{g}$ and $s^\prime_k = 1$ if $g_k \in G^{-}_{g}$. This state has signature $v$ with $v_k =\overline{o}$ for all $k \in \{1,\ldots ,n\}$. For all $f_g \in \mathcal{F}_{g}$, $f_{g} \preceq \emph{sup }\mathcal{F}_{g}$ and thus $f_{g}(\mathbf{s}^\prime) =1$.

\end{proof}

\begin{proof}[Proof of Proposition \ref{prop:incr_decr-trans}]

First of all, observe that the size of the set of true states of $f_g$ ($\mathbf{s}$ such that $f_{g}(\mathbf{s}) = 1$) grows as $f_{g}$ is localized upper in the HD of $(\mathcal{F}_{g},\preceq)$. Consequently, bounds for the numbers of increasing and decreasing transitions in $\EE_{g}$ are obtained for the top and bottom regulatory functions of $(\mathcal{F}_{g},\preceq)$.

For the case where $g$ is not auto-regulated ($g \notin G_g$), Proposition \ref{prop:transitions} states that the number of transitions in $\EE_{g}$ does not depend on $f_{g}$ and is equal to $2^{n-1}$. Moreover, from the proof of the proposition, we have that there is exactly one transition linking each pair of states $(s_1,\ldots, s_{n-1}, 0), (s_1,\ldots, s_{n-1}, 1)$, either an increasing transition if $f_{g}(s_1,\ldots, s_n) = 1$ or a decreasing transition if $f_{g}(s_1,\ldots, s_n) = 0$. Thus, when changing $f_{g}$, at most the orientation of the transition between such pair of states changes. In particular, for the top regulatory function (\emph{presence of at least one activator or absence of at least one inhibitor}), the only state for which $f_{g}(\mathbf{s}) = 0$ is the state specified in Lemma \ref{lem:true_false}. Thus $\mathcal{T}_{f_g}$ encompasses all but one ($2^{n-1} - 1$) increasing transitions. On the other hand, for the bottom regulatory function (\emph{presence of all activators and absence of all inhibitors}) the only state for which $f_{g}(\mathbf{s}) = 1$ is the one defined in Lemma \ref{lem:true_false}.
As a consequence, $\mathcal{T}_{f_g}$ contains only 1 increasing transition and $2^{n} - 1$ decreasing transitions.

Examining now the case where $g$ is positively auto-regulated ($g \in G^{+}_g$), Proposition \ref{prop:transitions} states that the number of transitions in $\EE_{g}$ is strictly lower than the number of transitions for a non auto-regulated component ($|\mathcal{T}_{f_g}| < 2^{(n - 1)}$). 
For $f_{g} =$\emph{sup }$\mathcal{F}_{g}$, the only state $\mathbf{s}$ for which $f_{g}(\mathbf{s}) = 0$ is the state specified in Lemma \ref{lem:true_false} such that $s_n = 0$. Thus, $\mathcal{T}_{f_g}$ encompasses $(2^{n-1}) - 1$ increasing transitions and no decreasing transitions. A similar reasoning for \emph{inf }$\mathcal{F}_{g}$ shows that, in this case, $\mathcal{T}_{f_g}$ encompasses $2^{(n-1)} - 1$ decreasing transitions and no increasing transitions.

Finally, when $g$ is negatively auto-regulated ($g \in G^{-}_g$), Proposition \ref{prop:transitions} states that $2^{(n-1)} < |\mathcal{T}_{f_g}| \leq 2^{n}$. For $f_{g} =\emph{ sup }\mathcal{F}_{g}$, $f_{g}(\mathbf{s}) = 0$ for only one state with $s_n = 1$ and thus $\mathcal{T}_{f_g}$ encompasses $2^{n-1}$ increasing and $1$ decreasing transitions.
Similarly, for $f_{g} =\emph{ inf }\mathcal{F}_{g}$, $\mathcal{T}_{f_g}$ encompasses $2^{(n-1)}$ decreasing and 1 increasing transitions.

\end{proof}

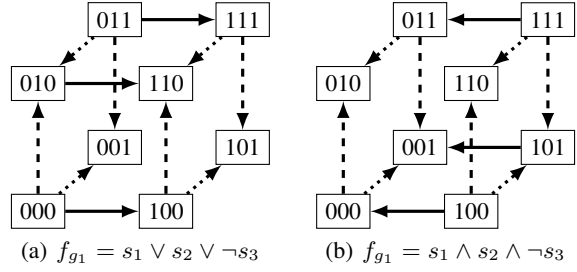
\begin{figure}
\begin{center}
\subfigure[$f_{g_1} = s_1 \vee s_2 \vee \neg s_3$]{
\begin{tikzpicture}
	\tikzstyle{stgstate} = [draw,font=\small]
	\tikzstyle{stgedge} = [-latex,very thick,font=\sffamily\normalsize\bfseries]

	\def\incx{1.7}
	\def\incy{1.7}
	\colorlet{darkgreen}{green!50!black}

	\node[stgstate] (000) at (0*\incx,0*\incy) {000};
	\node[stgstate] (100) at (1*\incx,0*\incy) {100};
	\node[stgstate] (001) at (0.6*\incx,0.5*\incy) {001};
	\node[stgstate] (101) at (1.6*\incx,0.5*\incy) {101};

	\node[stgstate] (010) at (0*\incx,1*\incy) {010};
	\node[stgstate] (110) at (1*\incx,1*\incy) {110};
	\node[stgstate] (011) at (0.6*\incx,1.5*\incy) {011};
	\node[stgstate] (111) at (1.6*\incx,1.5*\incy) {111};

	\draw[stgedge] (000) edge (100);
	\draw[stgedge] (010) edge (110);
	\draw[stgedge] (011) edge (111);
	\draw[stgedge,dashed] (000) edge (010);
	\draw[stgedge,dashed] (100) edge (110);
	\draw[stgedge,dashed] (011) edge (001);
	\draw[stgedge,dashed] (111) edge (101);
	\draw[stgedge,dotted] (000) edge (001);
	\draw[stgedge,dotted] (100) edge (101);
	\draw[stgedge,dotted] (011) edge (010);
	\draw[stgedge,dotted] (111) edge (110);
\end{tikzpicture}

\label{fig:stg-sup}}
\quad
\subfigure[$f_{g_1} = s_1 \wedge s_2 \wedge \neg s_3$]{
\begin{tikzpicture}
	\tikzstyle{stgstate} = [draw,font=\small]
	\tikzstyle{stgedge} = [-latex,very thick,font=\sffamily\normalsize\bfseries]

	\def\incx{1.7}
	\def\incy{1.7}
	\colorlet{darkgreen}{green!50!black}

	\node[stgstate] (000) at (0*\incx,0*\incy) {000};
	\node[stgstate] (100) at (1*\incx,0*\incy) {100};
	\node[stgstate] (001) at (0.6*\incx,0.5*\incy) {001};
	\node[stgstate] (101) at (1.6*\incx,0.5*\incy) {101};

	\node[stgstate] (010) at (0*\incx,1*\incy) {010};
	\node[stgstate] (110) at (1*\incx,1*\incy) {110};
	\node[stgstate] (011) at (0.6*\incx,1.5*\incy) {011};
	\node[stgstate] (111) at (1.6*\incx,1.5*\incy) {111};

	\draw[stgedge] (100) edge (000);
	\draw[stgedge] (101) edge (001);
	\draw[stgedge] (111) edge (011);
	\draw[stgedge,dashed] (000) edge (010);
	\draw[stgedge,dashed] (100) edge (110);
	\draw[stgedge,dashed] (011) edge (001);
	\draw[stgedge,dashed] (111) edge (101);
	\draw[stgedge,dotted] (000) edge (001);
	\draw[stgedge,dotted] (100) edge (101);
	\draw[stgedge,dotted] (011) edge (010);
	\draw[stgedge,dotted] (111) edge (110);
\end{tikzpicture}

\label{fig:stg-inf}}
\caption{STG of the regulatory graph of Figure \ref{fig:toyBN} for (a) $f_{g_1} =\emph{ sup }\mathcal{F}_{g_1}$ and (b) $f_{g_1} =\emph{ inf }\mathcal{F}_{g_1}$; in both cases $f_{g_2}(\mathbf{s}) = \neg s_3$ and $f_{g_3}(\mathbf{s}) = \neg s_2$.}
\label{fig:stg-sup-inf}
\end{center}
\end{figure}

Figure~\ref{fig:stg-sup-inf}(a) and \ref{fig:stg-sup-inf}(b) show the STG for $f_{g_1} =\emph{ sup }\mathcal{F}_{g_1}$ and for $f_{g_1} =\emph{ inf }\mathcal{F}_{g_1}$, considering the regulatory graph of Figure \ref{fig:toyBN}. In this example, $f_{g_1}$ is the only modifiable regulatory function, and since $g_1$ corresponds to a positive auto-regulated component, the contribution of $f_{g_1}$ to the transition structure of the STG can vary from $L_{g_1} = 0$ to $U_{g_1} = 3$ increasing/decreasing transitions. 

Observe that in the STG of Figure \ref{fig:toyBN} corresponding to $f_{g_1}(\mathbf{s}) = s_1 \vee (s_2 \land \neg s_3)$, where \emph{inf }$\mathcal{F}_{g_1}$ $\preceq f_{g_1} \preceq$ \emph{sup }$\mathcal{F}_{g_1}$, the number of increasing and decreasing transitions due to $f_{g_1}$ are $|\mathcal{T}^{+}_{f_{g_1}}| = 1$ and $|\mathcal{T}^{-}_{f_{g_1}}| = 0$, while upper and lower bounds of increasing and decreasing transitions are $U_{g_1} = 3$ and $L_{g_1} = 0$. 

\begin{figure*}
\centering
\begin{tabular}{cp{0.3cm}c}
5 regulators && 6 regulators\\
\begin{minipage}{0.45\textwidth}{{\bf A}\includegraphics[width=\textwidth]{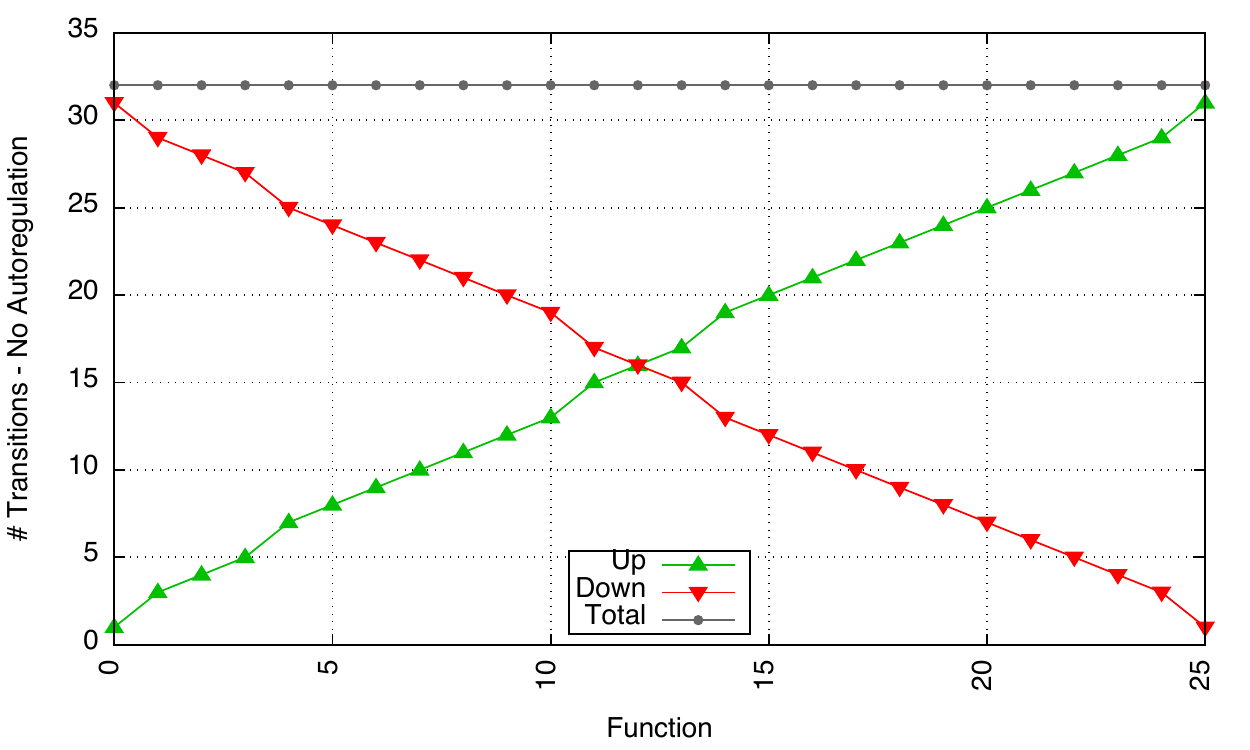}}\end{minipage}
&&\begin{minipage}{0.45\textwidth}{{\bf B}\includegraphics[width=\textwidth]{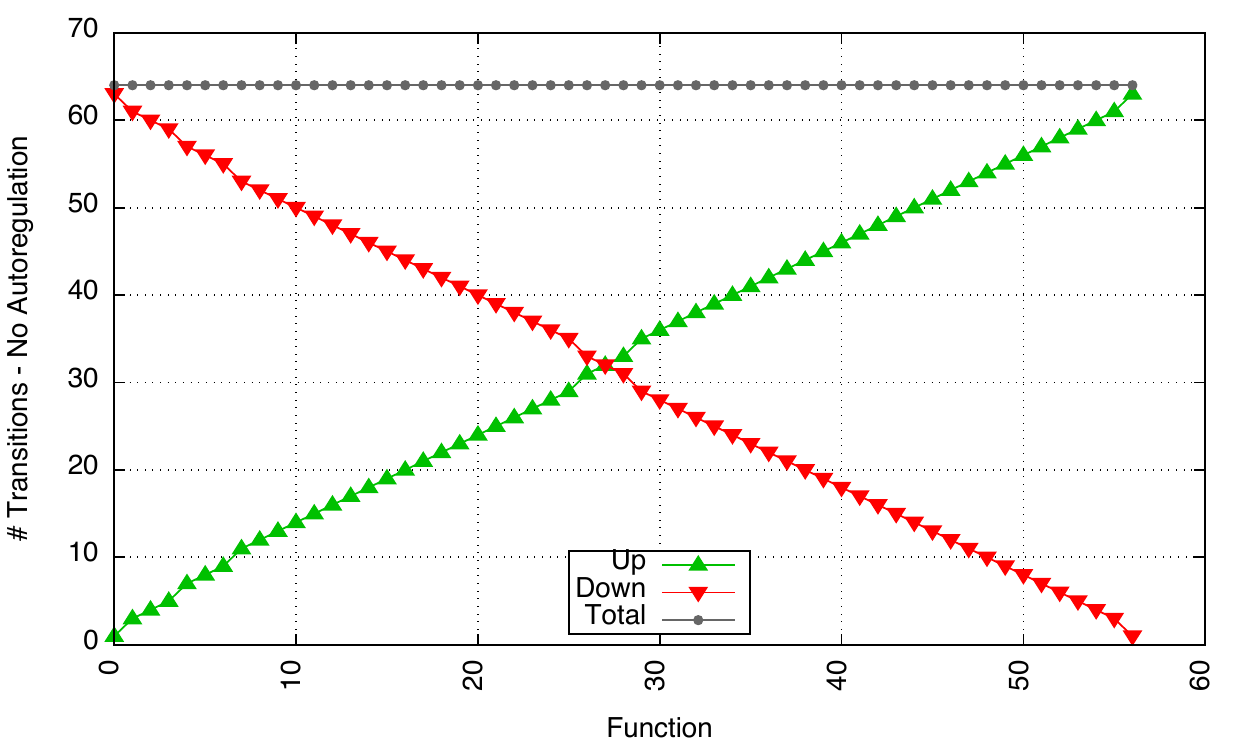}}\end{minipage}\\

\begin{minipage}{0.45\textwidth}{{\bf C}\includegraphics[width=\textwidth]{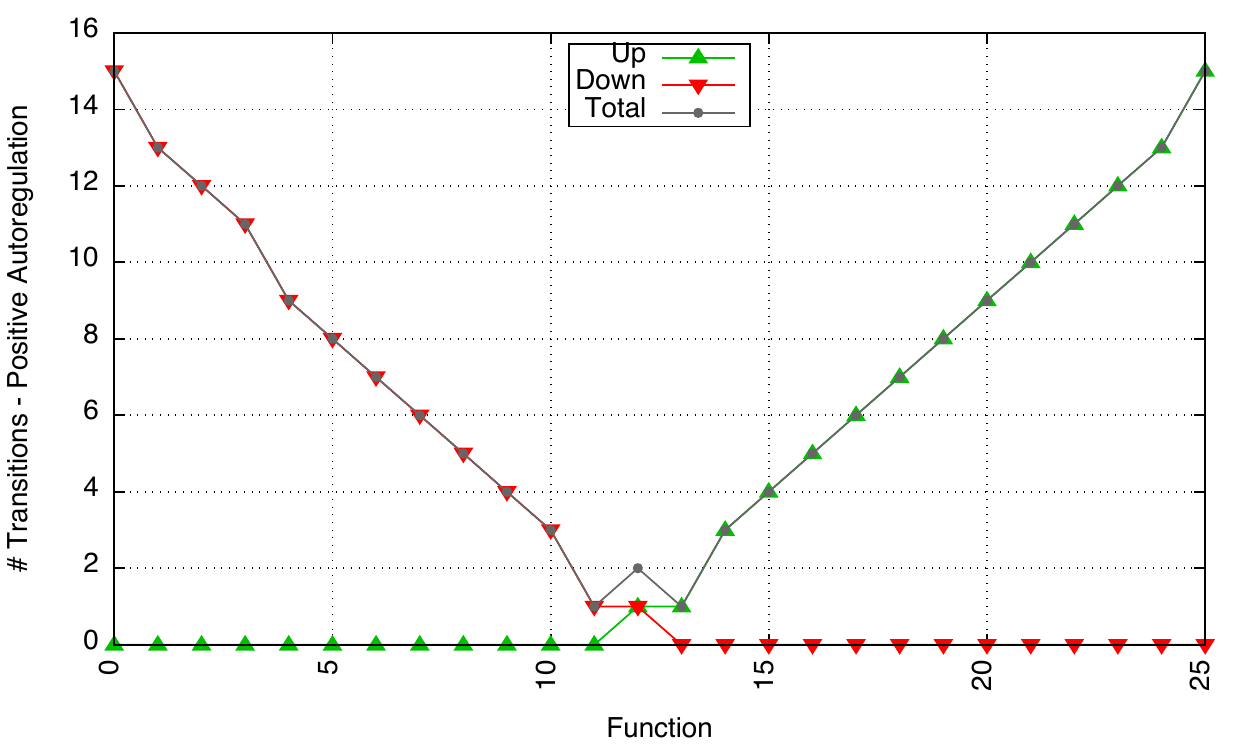}}\end{minipage}
&&\begin{minipage}{0.45\textwidth}{{\bf D}\includegraphics[width=\textwidth]{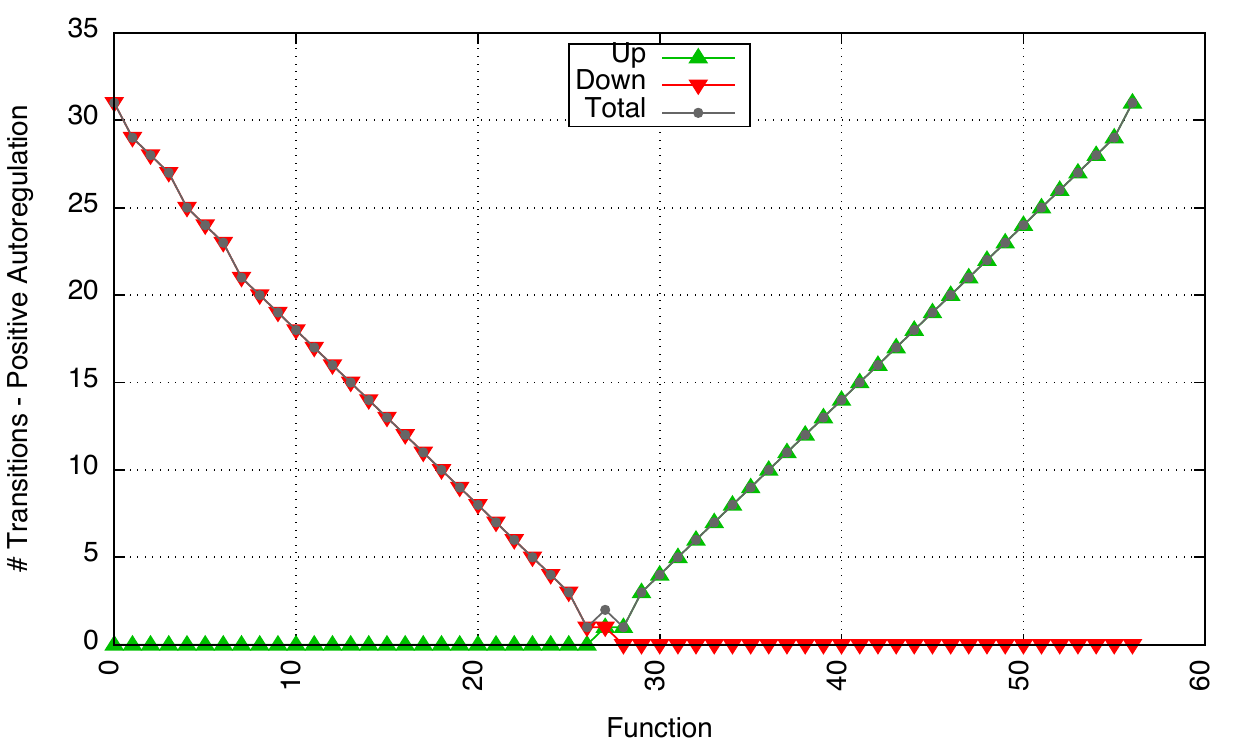}}\end{minipage}\\

\begin{minipage}{0.45\textwidth}{{\bf E}\includegraphics[width=\textwidth]{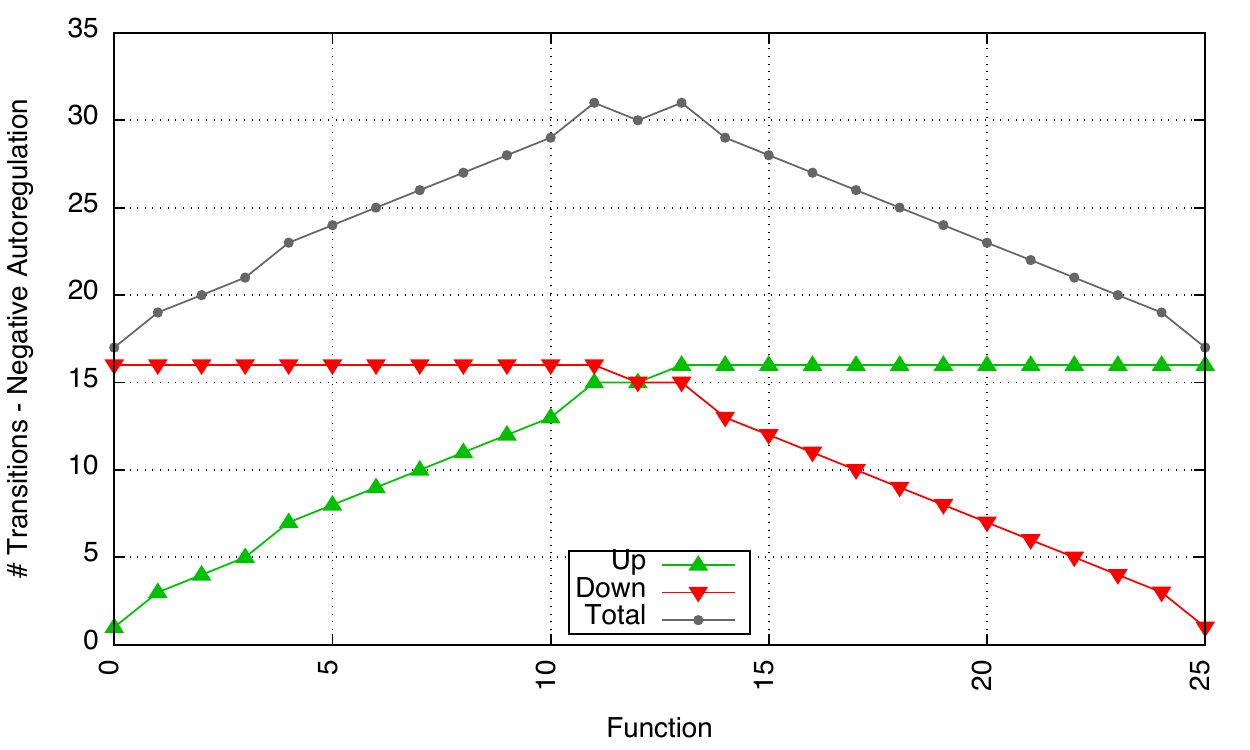}}\end{minipage}
&&\begin{minipage}{0.45\textwidth}{{\bf F}\includegraphics[width=\textwidth]{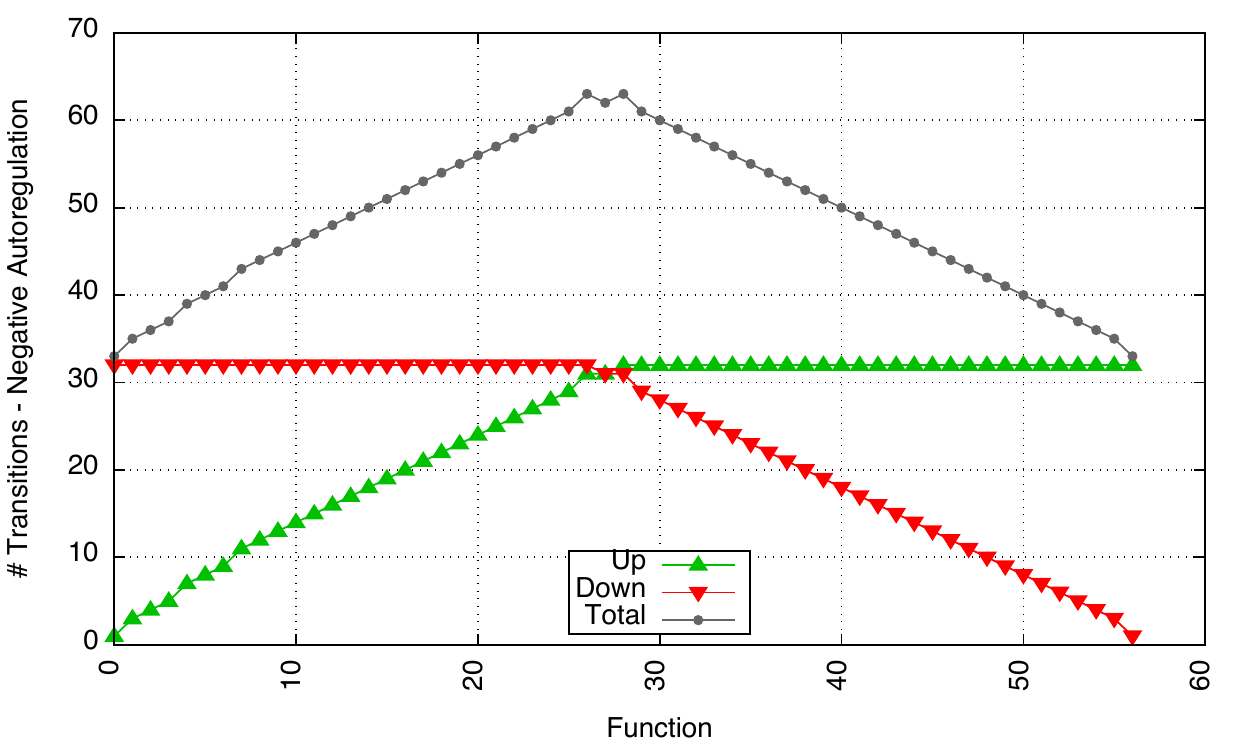}}\end{minipage}\\

\end{tabular}
\begin{caption}{Evolution of the number of transitions affecting a component regulated by 5 (left) or 6 (right) regulators, along a random path of the HD (from $\emph{inf }\mathcal{F}_{g}$ to $\emph{sup }\mathcal{F}_{g}$): A-B the component is not auto-regulated, {\it i.e.}, the Boolean functions are independent of its state; C-D the component activates itself, {\it i.e.}, the Boolean functions depend on the corresponding positive literal; E-F the component inhibits itself, {\it i.e.}, the Boolean functions depend on the corresponding negative literal. \label{fig:path_HD_transitions}}
\end{caption}
\end{figure*}

Thanks to the \emph{Rules to compute parents}, it is possible to circulate along paths in the HD of $(\mathcal{F}_g,\preceq)$, between $\emph{inf }\mathcal{F}_{g}$ and $\emph{sup }\mathcal{F}_{g}$, and assess the evolution of the numbers of transitions of each $f_g$ along those paths ({\it i.e.,} by varying regulatory functions).
Figure~\ref{fig:path_HD_transitions} illustrates such variations for components with five and six regulators, considering the cases of non auto-regulated and auto-regulated components.

\subsection{Special reference regulatory functions}



Here, we identify some Boolean regulatory functions $f_g$ in $(\mathcal{F}_{g},\preceq)$ that lead to specific relationships between $|\mathcal{T}^{+}_{f_g}|$ and $|\mathcal{T}^{-}_{f_g}|$, or to maximal or minimal total number of transitions in $\EE_g$.

\begin{prop}\label{prop:inc-dec-noautoreg}

Let $G_g$ be the set of regulators of a non auto-regulated component $g$ ($g \notin G_g$). For any $f_g \in \mathcal{F}_{g}$ we have:
\begin{enumerate}

\item $|\mathcal{T}^{+}_{f_g}| = |{\mathbb{T}}(f_g)|$; and
\item $|\mathcal{T}^{-}_{f_g}| = 2^{n} - |\mathbb{T}(f_g)|$.

\end{enumerate}
\end{prop}

\begin{proof}

The result straightforwardly follows from the proof of Proposition \ref{prop:transitions}: there is exactly one transition between any pair of states $(s_1,\ldots, s_{n-1}, 0),(s_1,\ldots, s_{n-1}, 1)$ that is an increasing transition if $f_{g}(s_1,\ldots, s_n) = 1$, or a decreasing transition if $f_{g}(s_1,\ldots, s_n) = 0$.

\end{proof}

\begin{cor}\label{prop:ratio_inc-dec}

Let $G_g$ be the set of regulators of a non auto-regulated component $g$ ($g \notin G_g$). If $f_g \in \mathcal{F}_{g}$ is such that $|\mathbb{T}(f_g)| = 2^{n-1}$, then
\begin{enumerate}

\item $|\mathcal{T}^{+}_{f_g}| = |\mathcal{T}^{-}_{f_g}| = 2^{n-1}$;
\item for all $f^\prime_g \in \mathcal{F}_{g}$,

$f^\prime_g \preceq f_g \implies |\mathcal{T}^{-}_{f^\prime_g}| \ge |\mathcal{T}^{+}_{f^\prime_g}|$

$f_g \preceq f^\prime_g \implies |\mathcal{T}^{-}_{f^\prime_g}| \le |\mathcal{T}^{+}_{f^\prime_g}|$
\end{enumerate}
\end{cor}

\begin{proof}

The first item follows from Proposition \ref{prop:inc-dec-noautoreg}. The second items directly follows from the fact that $|\mathcal{T}^{+}_{f_g}|$ increases and $|\mathcal{T}^{-}_{f_g}|$ decreases as $f_g$ becomes greater in $(\mathcal{F}_{g},\preceq)$.

\end{proof}


\begin{prop}\label{prop:max-min_trans}

Let $G_g$ be the set of regulators of an auto-regulated component $g$ ($g \in G_g$, and thus $|G_g| > 1$). For the Boolean regulatory function 
\begin{equation}\label{eq:f*}
f^{*}_g = \bigvee_{\substack{g_k \in G^{+}_{g} \\ k \ne n}} (u \land s_k) \bigvee_{\substack{g_k \in G^{-}_{g} \\ k \ne n}} (u \land \neg s_k),\end{equation}

\begin{enumerate}
\item if $g \in G^{-}_g$, with $u = \neg s_n$, then 
\begin{enumerate}
\item $|\mathcal{T}^{-}_{f^{*}_g}| = 2^{n-1}$, $|\mathcal{T}^{+}_{f^{*}_g}| = 2^{n-1} - 1$, and thus $|\mathcal{T}_{f^{*}_g}| = 2^{n} -1$;
\item for all $f_g \in \mathcal{F}_{g}$,

$f_g \preceq f^{*}_g \implies |\mathcal{T}^{-}_{f_g}| = 2^{n-1}, |\mathcal{T}^{+}_{f_g}| \le 2^{n-1} - 1$

$f^{*}_g \preceq f_g$ $\implies$ $|\mathcal{T}^{-}_{f_g}| \le 2^{n-1}$, $|\mathcal{T}^{+}_{f_g}| = 2^{n-1} - 1$;

\end{enumerate}
\item if $g \in G^{+}_g$, with $u = s_n$, then 
\begin{enumerate}
\item $|\mathcal{T}_{f^{*}_g}| = |\mathcal{T}^{-}_{f^{*}_g}| = 1$;
\item for all $f_g \in \mathcal{F}_{g}$, 

$f_g \preceq f^{*}_g$ $\implies$ $|\mathcal{T}^{-}_{f_g}| \ge 1$, $|\mathcal{T}^{+}_{f_g}| = 0$

$f^{*}_g \preceq f_g$ $\implies$ $|\mathcal{T}^{-}_{f_g}| \le 1$, $|\mathcal{T}^{+}_{f_g}| \ge 0$.

\end{enumerate}
\end{enumerate}
\end{prop}

\begin{proof}

Let us first consider the case where $g \in G^{-}_g$. Then $f^{*}_g(\mathbf{s}) = 0$ for all states $\mathbf{s}\in \BB^n$ such that $s_n = 1$ because all the clauses in Equation \ref{eq:f*} contain $ \neg s_n$. Therefore, in the STG $\EE_{g}$, there is a decreasing transition going out each of those $2^{n-1}$ states. Moreover, for the states such that $s_n = 0$, the only state for which $f^{*}_g(\mathbf{s}) = 0$ is when all the activators are absent ($s_k = 0$ for $s_k \in G^{+}_{g}$) and all the inhibitors but $g$ are present ($s_k = 1$ for $g_k \in G^{-}_{g}$, $k \ne n$). In other words, $f^{*}_g(\mathbf{s}) = 1$ in all but one state for which $s_n = 0$. Therefore, in the STG $\EE_{g}$, there is an increasing transition going out each of those $2^{n-1}-1$ states. The total number of transitions is thus $|\mathcal{T}_{f^{*}_g}| = |\mathcal{T}^{+}_{f^{*}_g}| + |\mathcal{T}^{-}_{f^{*}_g}| = 2^{n} - 1$.

The case (1b) follows from the facts that $|\mathcal{T}^{+}_{f_g}|$ increases and $|\mathcal{T}^{-}_{f_g}|$ decreases when $f_g$ becomes greater in $(\mathcal{F}_{g},\preceq)$, that $|\mathcal{T}^{-}_{f^{*}_g}| = 2^{n-1} = U_g$ and that, if $|G_g| > 1$, the upper bound for $|\mathcal{T}_{f_g}|$ is $2^{n-1} - 1$. 

If $g \in G^{+}_g$, a similar reasoning shows that there are $2^{n-1}$ states $\mathbf{s} \in \BB^n$ for which $f^{*}_g(\mathbf{s}) = 0$ and $s_n =0$, and only one state in which $f^{*}_g(\mathbf{s}) = 0$ and $s_n =1$. This implies that there is a single (decreasing) transition in $\EE_{g}$, {\it i.e,} $|T_{f_g}| = |\mathcal{T}^{-}_{f_g}| = 1$; (2b) also follows from arguments similar to those employed for (1b).

\end{proof}

The numbers of transitions in Proposition \ref{prop:max-min_trans} correspond to the maximal (resp. minimal) numbers reached for the case of a component negatively (resp. positively) auto-regulated, and with multiple regulators. Those numbers are obtained for the functions defining maximally functional auto-regulation. 
These functions enounce that the auto-regulated component $g$ is activated in \emph{the absence of $g$ and the presence of at least one other inhibitor, or in the absence of $g$ and the presence of at least one activator}, in the case of an inhibitory auto-regulation, and in \emph{the presence of $g$ and of at least one other activator, or the presence of $g$ and the absence of at least one inhibitor}, in the case of an activatory auto-regulation.

\subsection{Levels of Boolean regulatory functions in the PO-Set}\label{sec:level}

In order to qualitatively evaluate the level of a particular regulatory function in the {PO-Set} $(\mathcal{F}_{g},\preceq)$ it is important to define a measure of its distance to the boundary functions. In this sense, an index associated to any regulatory function $f \in \mathcal{F}_{g}$ is introduced in what follows.

\begin{defin}\label{def:level}

Let $\mathcal{R}$ be a Boolean network with $n$ components regulating $g \in G$, and let $f_g = C_1 \vee \ldots \vee C_m$ the CDNF representation of the regulatory function of $g$, in which the clauses are ordered so that, if $l_k$ denotes the dimension of the subspace of the clause $C_k$, $l_k \geq l_j$ for $k < j$. The \emph{level} $l(f_g)$ of $f_g$ is defined as the ordered $m$-tuple $(l_1(f_g),\dots,l_m(f_g))$.

\end{defin}

The level specified in Definition \ref{def:level} associates to a regulatory function, the list of dimensions of the subspaces of its clauses in a decreasing order. 

Note that for a {PO-Set} $(\mathcal{F}_{g},\preceq)$ on $\{1,\ldots, n\}$, $l\textrm{(\emph{sup }}\mathcal{F}_g) = \underbrace{(n-1,\ldots,n-1)}_{n \textrm{ times}}$, and $l($\emph{inf }$\mathcal{F}_g) = (0)$. 

In the example of the {PO-Set} corresponding to $(\mathcal{F}_{g_1},\preceq)$ for the Boolean network of Figure \ref{fig:toyBN}, $l\textrm{(\emph{sup }}\mathcal{F}_{g_1}) = (2,2,2)$, $l\textrm{(\emph{inf }}\mathcal{F}_{g_1}) = (0)$, and $l(s_1 \vee (s_2 \land \neg s_3)) = (2,1)$. 

A total order $\leq$ can be defined on $L_{\mathcal{F}_{g}}$, set of the levels of the functions in $(\mathcal{F}_{g},\preceq)$ as follows: given $f, f^\prime \in \mathcal{F}_{g}$ such that $f = C_1 \vee \ldots \vee C_m$ and $f^\prime = C^\prime_1 \vee \ldots \vee C^\prime_{m^\prime}$, $l(f) \leq l(f^\prime)$ if and only if one of the following conditions holds:
\begin{itemize}
\item[(i)] there exists $k \in \{1,\ldots,min(m,m^\prime)\}$ for which $l_k(f) < l_k(f^\prime)$, or 
\item[(ii)] $l_k(f) = l_k(f^\prime)$ for all $k \in \{1,\ldots,m\}$ and $m \leq m^\prime$.
\end{itemize}

It is straightforward to verify that given the {PO-Set} $(\mathcal{F}_{g},\preceq)$ on $\{1,\ldots, n\}$, for any $f \in \mathcal{F}_{g}$, $l\textrm{(\emph{inf }}\mathcal{F}_{g}) \leq l(f) \leq l\textrm{(\emph{sup }}\mathcal{F}_g)$. The following proposition generalizes this relationship.

\begin{prop}\label{prop:level}

For $f, f^\prime \in \mathcal{F}_{g}$, if $f \preceq f^\prime$ then $l(f) \leq l(f^\prime)$. 

\end{prop}

\begin{proof}

 It follows from Equation \ref{eq:two_relations} that $f \preceq f^\prime$ implies $S \preceq S^\prime$, where $S$ and $S^\prime$ are the set-representations of $f$ and $f^\prime$ respectively. From the definition of $S$, and after an appropriate ordering of the elements in $S$, we have that, for each clause $C_k$ of $f$, there exists a $\sigma_k \in S$ such that $l_k(f) = n - |\sigma_{k}|$. The same applies to clauses of $f^\prime$ and elements of $S^\prime$. 
Now, from the definition of $\preceq$ on $\mathcal{S}$, $\forall \sigma \in S\text{, } \exists \sigma^{\prime} \in S^{\prime} \text{ such that } \sigma \supseteq \sigma^{\prime}$, which in turn implies $|\sigma| \ge |\sigma^{\prime}|$ and thus $n - |\sigma| \le n - |\sigma^{\prime}|$. The definition of the total order $\le$ on $L_{\mathcal{F}_g}$ does the rest.

\end{proof}

\begin{figure}[!tpb]
	\centering
	\resizebox{0.5\textwidth}{!}{%
\begin{tikzpicture}
  \tikzstyle{hdstate} = [ellipse,draw,blue,font=\bfseries\small]
	\tikzstyle{hdedge} = [gray,font=\sffamily\normalsize\bfseries]
	\tikzstyle{hdfunc} = [red,font=\large]

	\def\xspace{3.5}
	\def\yspace{2}
	
	\node[hdstate] (1-2-3) at (0,4*\yspace) {\{\{1\},\{2\},\{3\}\}};

	\node[hdstate] (3-12) at (-1*\xspace,3*\yspace) {\{\{3\},\{1,2\}\}};
	\node[hdstate] (2-13) at (0,3*\yspace) {\{\{2\},\{1,3\}\}};
	\node[hdstate] (1-23) at (1*\xspace,3*\yspace) {\{\{1\},\{2,3\}\}};

	\node[hdstate] (12-13-23) at (0,2*\yspace) {\{\{1,2\},\{1,3\},\{2,3\}\}};

	\node[hdstate] (12-23) at (-1*\xspace,1*\yspace) {\{\{1,2\},\{2,3\}\}};
	\node[hdstate] (12-13) at (0,1*\yspace) {\{\{1,2\},\{1,3\}\}};
	\node[hdstate] (23-13) at (1*\xspace,1*\yspace) {\{\{1,3\},\{2,3\}\}};

	\node[hdstate] (123) at (0,0) {\{\{1,2,3\}\}};

	\draw[hdedge] (3-12) edge node[black] {} (1-2-3);
	\draw[hdedge] (2-13) edge node[black] {} (1-2-3);
	\draw[hdedge] (1-23) edge node[black] {} (1-2-3);

	\draw[hdedge] (12-13-23) edge node[black] {} (3-12);
	\draw[hdedge] (12-13-23) edge node[black] {} (2-13);
	\draw[hdedge] (12-13-23) edge node[black] {} (1-23);

	\draw[hdedge] (12-23) edge node[black] {} (12-13-23);
	\draw[hdedge] (12-13) edge node[black] {} (12-13-23);
	\draw[hdedge] (23-13) edge node[black] {} (12-13-23);

	\draw[hdedge] (123) edge node[black] {} (12-23);
	\draw[hdedge] (123) edge node[black] {} (12-13);
	\draw[hdedge] (123) edge node[black] {} (23-13);

	\path[hdfunc] (1-2-3) ++(2.5,0) node {(2,2,2)};
	\path[hdfunc] (123) ++(1.7,0) node {(0)};
	\path[hdfunc] (1-23) ++(2.1,0) node {(2,1)};
	\path[hdfunc] (12-13-23) ++(3.2,0) node {(1,1,1)};
	\path[hdfunc] (23-13) ++(2.3,0) node {(1,1)};
\end{tikzpicture}
	}%
	\caption{Levels of the regulatory functions of $g_1$ from the Boolean network of Figure \ref{fig:toyBN} (and more generally of any component with 3 regulators). The same level applies to all functions in the same layer of the Hasse Diagram.}\label{fig:HD_L}
\end{figure}
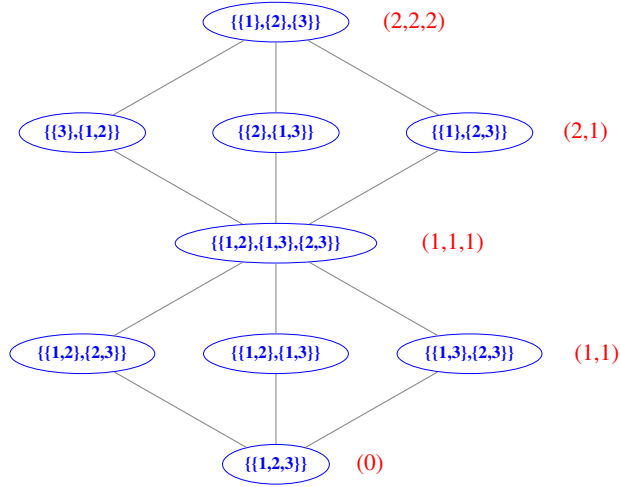

Figure \ref{fig:HD_L} illustrates the levels of the regulatory functions in $\mathcal{F}_{g_1}$ for the Boolean network of Figure \ref{fig:toyBN}. It is clear from the definition that these levels depend only on the set-representations of the functions and not on the signs of the regulatory interactions.

The function levels in the {PO-Set} $(\mathcal{F}_{g},\preceq)$ provide a measure of the distances to the boundary functions, and consequently a measure of the impact on the dynamics of the corresponding Boolean network.

\section{Applications}\label{sec:applications}

\subsection{Assessing some common regulatory functions}
We first consider a particular case of \emph{Majority Rule} (${MR}_g$), a specific type of threshold functions \cite{majorityrule}. This $MR_g$ function is stated in \cite{majorityrule} as an inequality that corresponds to \emph{the difference between number of present activators plus absent inhibitors and number of absent activators plus present inhibitors is greater or equal zero}. In general, equality is evaluated apart, with an associated probability $p_g$. For the case $p_g = 1$ (equality always accepted) the function can be translated as \emph{number of present activators plus absent inhibitors is at least $r = \lceil{n/2}\rceil$}. More generally, we could consider the case where $1 \leq r \leq n$. This $MR_g$ can be written in the \emph{MDNF} as $$MR_g = \bigvee_{i \in \{1,\ldots, {n \choose r}\}} C_j$$ with $|E_j| = r$, for all $j \in \{1,\ldots, {n \choose r}\}$. The level of the $MR_g$ function is then $l(MR_g) = \underbrace{(n-r,\ldots,n-r)}_{{n \choose r} \textrm{ times}}$.


$MR_g$ function is such that the greater the threshold $r$ is, the lower the level of the function, the limits being exactly \emph{sup }$\mathcal{F}_g$ (for $r = 1$) and \emph{inf }$\mathcal{F}_g$ (for $r = n$). The special threshold case considered in \cite{majorityrule} ($r = \lceil{n/2}\rceil$) for $p_g = 1$ can be equivalently stated as \emph{the number of present activators plus absent inhibitors is at least the same as the number of absent activators plus present inhibitors}. For instance, in the case where $n=3$, $MR_g$   set-representation is $S_g = \{\{1,2\},\{1,3\},\{2,3\}\}$ with level $l(MR_g) = (1,1,1)$.

Another regulatory function of interest, when $G^{+}_g \ne \emptyset$, is the one  stated as \emph{presence of at least one activator and absence of all inhibitors}. This function denoted here as $NI_g$ (No Inhibitors), can be represented in its \emph{MDNF} as $$NI_g = \bigvee_{j \in \{1,\ldots, |G^{+}_g|\}} C_j$$ with $|E_j| = |G^{-}_g|+1$ and such that $k \in E_j$ for all $g_k \in G^{-}_g$ and for one $g_k \in G^{+}_g$, $j \in \{1,\ldots, |G^{+}_g|\}$. The level of $NI_g$ is $l(NI_g) = \underbrace{(n-(|G^{-}_g|+1),\ldots,n-(|G^{-}_g|+1))}_{|G^{+}_g| \textrm{ times}}$. In the case of the example of Figure \ref{fig:toyBN}, $NI_{g_1} = (q_1 \land \neg q_3) \vee (q_2 \land \neg q_3)$ and $l(NI_{g_1}) = (1,1)$. For a  fixed number of regulators $|G_g|$, the greater the number of \emph{inhibitors}, the lower the level of the $NI_g$ function. The upper limiting level for $NI_g$ is \emph{sup }$\mathcal{F}_g$ when there is no inhibitor in the set of regulators ($G^{-}_g = \emptyset$); the lowest possible level is for $NI_g = \textrm{\emph{inf }}\mathcal{F}_g$ when all but one regulatory components are inhibitors.

\subsection{Stochasticity in Boolean networks}\label{sec:stochastic}

In this section, we explore the use of the previous results to assess robustness of Boolean Networks (BN) by adding some stochasticity in the regulatory functions. 

To introduce stochasticity in Boolean Networks (BN), several authors considered associating ensembles of Boolean functions to the model components with a probabilistic selection of one function at each simulation step \cite{garg2009,shmulevich2002}. Furthermore, robustness of Boolean networks has been investigated by perturbating the functions of the components \cite{xiao2007}. 

Here, we consider the Probabilistic Boolean Networks (PBN) as introduced by Shmulevich {\it et al.} \cite{shmulevich2002}, where each node is associated with a set of regulatory functions (at least one), each being attributed a probability. Note however that these functions can be any Boolean function, including degenerate and non-monotone functions. At each simulation step, a function is chosen for each component, and appropriate variable updates are performed synchronously to get the successor of the current state. In a PBN $\RR=(G,R,\FFF)$, where now $\FFF$ is a set $\{F_i=\{(f_i^k,p_i^k)\}_{k=1,\dots,|F_i|}\}_{i=1,\dots, |G|}$, where each component $g_i$ is associated with a set of $k$ Boolean regulatory functions, each associated with a probability $p_i^k$. At each step, the number of realizations of the PBN is $\Pi_{i=1,\dots,|G|}|F_i|$. 

%

Here, we perform the simulation of these networks using the software tool BoolNet \cite{boolnet}.
The local search of the set of regulatory Boolean functions to revisit the experiments proposed in \cite{garg2009}, is illustrated with the model of T helper cell differentiation from Mendoza \& Xenarios \cite{mendoza2006}. For this model, Table \ref{table:Thfunctions} provides the reference functions as well as their neighbors. 

\begin{table*}[t]
  \centering
 \resizebox{0.98\textwidth}{!}{\begin{tabular}{c|c|c|p{5.3cm}|p{8.6cm}}
Node & NbReg & NbFun. & \multicolumn{1}{c|}{Reference Function}&\multicolumn{1}{c}{Neighbouring Functions}\\\hline
$\mbox{GATA3}$&3&9&$(\neg \mbox{Tbet} \!\wedge\! \mbox{STAT6})\!\vee\!(\neg \mbox{Tbet} \!\wedge\! \mbox{GATA3})$&$\neg \mbox{Tbet} \!\wedge\! \mbox{STAT6}\!\wedge\! \mbox{GATA3}$\\\cline{5-5}
&&&&$(\neg \mbox{Tbet} \!\wedge\! \mbox{STAT6})\!\vee\!(\neg \mbox{Tbet} \!\wedge\! \mbox{GATA3})\!\vee\! (\mbox{STAT6}\!\wedge\! \mbox{GATA3})$\\\cline{5-5}
&&&&$(\neg \mbox{Tbet} \!\wedge\! \mbox{STAT6})\!\vee\! (\mbox{STAT6}\!\wedge\! \mbox{GATA3})^\ast$\\\cline{5-5}
&&&&$(\neg \mbox{Tbet} \!\wedge\! \mbox{GATA3})\!\vee\! (\mbox{STAT6}\!\wedge\! \mbox{GATA3})^\ast$\\\hline

$\mbox{\mbox{IFNb}R}$	&1&1&$\mbox{IFNb}$&\mbox{ }\\\hline

	 $\mbox{IFNg}$&5&	6894&$(\neg \mbox{STAT3} \!\wedge\!  \mbox{NFAT})\!\vee\!  (\neg \mbox{STAT3} \!\wedge\! \mbox{Tbet}) \!\vee\!  (\neg \mbox{STAT3} \!\wedge\! \mbox{IRAK})\!\vee\! (\neg \mbox{STAT3}  \!\wedge\! \mbox{NFAT})  $
	 & $(\neg \mbox{STAT3} \!\wedge\! \mbox{IRAK}) \!\vee\!
	    (\neg \mbox{STAT3} \!\wedge\! \mbox{NFAT}) \!\vee\! 
	    (\neg \mbox{STAT3} \!\wedge\! \mbox{Tbet}) \!\vee\!
			(\neg \mbox{STAT3} \!\wedge\! \mbox{NFAT})$\\\cline{5-5}
	 & & & & $(\neg \mbox{STAT3} \!\wedge\! \mbox{IRAK} \!\wedge\! \mbox{Tbet}) \!\vee\!
	          (\neg \mbox{STAT3} \!\wedge\! \mbox{NFAT}) \!\vee\!
						(\neg \mbox{STAT3} \!\wedge\! \mbox{STAT4})$\\\cline{5-5}
	 & & & & $(\neg \mbox{STAT3} \!\wedge\! \mbox{Tbet}) \!\vee\!
	          (\neg \mbox{STAT3} \!\wedge\! \mbox{IRAK} \!\wedge\! \mbox{NFAT}) \!\vee\!
						(\neg \mbox{STAT3} \!\wedge\! \mbox{STAT4})$\\\cline{5-5}
   & & & & $(\neg \mbox{STAT3} \!\wedge\! \mbox{Tbet}) \!\vee\!
	          (\neg \mbox{STAT3} \!\wedge\! \mbox{NFAT}) \!\vee\!
						(\neg \mbox{STAT3} \!\wedge\! \mbox{IRAK} \!\wedge\! \mbox{STAT4})$\\\cline{5-5}
	 & & & & $(\neg \mbox{STAT3} \!\wedge\! \mbox{IRAK} ) \!\vee\!
	          (\neg \mbox{STAT3} \!\wedge\! \mbox{NFAT} \!\wedge\! \mbox{Tbet}) \!\vee\!
						(\neg \mbox{STAT3} \!\wedge\! \mbox{STAT4})$\\\cline{5-5}
	 & & & & $(\neg \mbox{STAT3} \!\wedge\! \mbox{IRAK} ) \!\vee\!
	          (\neg \mbox{STAT3} \!\wedge\! \mbox{NFAT}) \!\vee\!
						(\neg \mbox{STAT3} \!\wedge\! \mbox{STAT4} \!\wedge\! \mbox{Tbet})$\\\cline{5-5}
	 & & & & $(\neg \mbox{STAT3} \!\wedge\! \mbox{IRAK} ) \!\vee\!
	          (\neg \mbox{STAT3} \!\wedge\! \mbox{Tbet}) \!\vee\!
						(\neg \mbox{STAT3} \!\wedge\! \mbox{NFAT} \!\wedge\! \mbox{STAT4})$\\\cline{5-5}
	 & & & & Plus 10 sibling functions$^{*}$ \\\hline

$\mbox{IFNgR}$&	1&1&	\mbox{IFNg}&\mbox{ }\\\hline
$\mbox{IL10}$&	1&1&	\mbox{GATA3}&\mbox{ }\\\hline
\mbox{IL10R}&	1&1&	\mbox{IL10}&\mbox{ }\\\hline

\mbox{\mbox{IL12}R}&	2&2&	$\neg \mbox{STAT6} \!\wedge\! \mbox{IL12}$&$\neg \mbox{STAT6} \!\vee\! \mbox{IL12}$\\\hline
\mbox{\mbox{IL18}R}&	2&2&	$\neg \mbox{STAT6} \!\wedge\! \mbox{IL18}$&$\neg \mbox{STAT6} \!\vee\! \mbox{IL18}$\\\hline

\mbox{IL4}&2&	2&	$\mbox{GATA3} \!\wedge\! \neg \mbox{STAT1}$&	$\mbox{GATA3} \!\vee\! \neg \mbox{STAT1}$\\\hline
\mbox{IL4}R	&2& 2	&$\mbox{IL4} \!\wedge\! \neg \mbox{SOCS1}$& $\mbox{IL4} \!\vee\! \neg \mbox{SOCS1}$\\\hline
\mbox{IRAK}	&1&1&	\mbox{\mbox{IL18}R}&\mbox{ }\\\hline
\mbox{JAK1}&2&	2	&$\mbox{IFNgR} \!\wedge\! \neg \mbox{SOCS1}$&$\mbox{IFNgR} \!\vee\! \neg \mbox{SOCS1}$\\\hline

\mbox{NFAT}&1&	1	&\mbox{TCR}&\mbox{ }\\\hline

\mbox{SOCS1}&2&	2&$\mbox{STAT1} \!\vee\! \mbox{Tbet}$&$\mbox{STAT1} \!\wedge\! \mbox{Tbet}$\\\hline
\mbox{STAT1}&2&	2&$\mbox{JAK1} \!\vee\! \mbox{\mbox{IFNb}R}$&$\mbox{JAK1} \!\wedge\! \mbox{\mbox{IFNb}R}$\\\hline

$\mbox{STAT3}$	&1&1&	$\mbox{IL10R}$ &\mbox{ }\\\hline
\mbox{STAT4}&2&	2&$\neg \mbox{GATA3} \!\wedge\! \mbox{\mbox{IL12}R}$&$\neg \mbox{GATA3} \!\vee\! \mbox{\mbox{IL12}R}$\\\hline
\mbox{STAT6}&1&	1&	$\mbox{IL4R}$&\mbox{ }\\\hline

$\mbox{Tbet}$&	3&9&$(\neg \mbox{GATA3} \!\wedge\! \mbox{STAT1})\!\vee\! (\neg \mbox{GATA3} \!\wedge\! \mbox{Tbet})$&$\neg \mbox{GATA3} \!\wedge\! \mbox{STAT1}\!\wedge\! \mbox{Tbet}$\\\cline{5-5}
&&&&$(\neg \mbox{GATA3} \!\wedge\! \mbox{STAT1})\!\vee\!(\neg \mbox{GATA3} \!\wedge\! \mbox{Tbet})\!\vee\! (\mbox{STAT1}\!\wedge\! \mbox{Tbet})$\\\cline{5-5}
&&&&$(\neg \mbox{GATA3} \!\wedge\! \mbox{STAT1})\!\vee\! (\mbox{STAT1}\!\wedge\! \mbox{Tbet})^\ast$\\\cline{5-5}
&&&&$(\neg \mbox{GATA3} \!\wedge\! \mbox{Tbet})\!\vee\! (\mbox{STAT1}\!\wedge\! \mbox{Tbet})^\ast$\\\hline

\mbox{IFNb} & 0 & 1 & False&\mbox{ }\\\hline
\mbox{IL12} & 0 & 1 & False&\mbox{ }\\\hline
\mbox{IL18} & 0 & 1 & False&\mbox{ }\\\hline
\mbox{TCR}  & 0 & 1 & False&\mbox{ }\\\hline
\end{tabular}}
\caption{\label{table:Thfunctions}Boolean model of the Mendoza \& Xenarios' T helper cell regulatory network \cite{mendoza2006} indicating, for each node, the number of regulators (2nd column), the number of compliant functions (3rd column), its original reference function (4th column) and the neighboring functions (last column). We consider direct parents and children, as well as siblings, {\it i.e.}, functions that share the same direct parents or children (indicated $^\ast$).}
\end{table*}

Considering that the reference function is indeed chosen (or effective) with probability $0.8$, we first start by distributing the remaining probability to the direct parent/child functions.
Doing so for a single component, allows to assess the \emph{criticality} of certain components, {\it e.g.} the function of IL4 is essential to maintain the expected behavior (differentiation to Th1, possibly with some cells maintaining a Th0).

\begin{figure*}[bt!]
  \centering
 \resizebox{0.98\textwidth}{!}{
\begin{tabular}{ll}
	\multicolumn{1}{c}{\begin{minipage}{0.52\textwidth}\includegraphics[width=\textwidth]{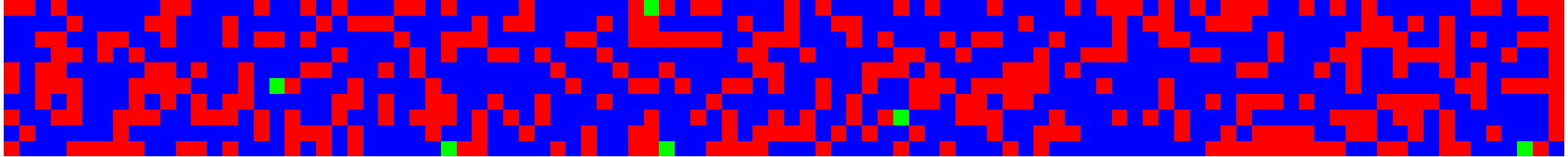}\end{minipage}} 
&
	\multicolumn{1}{c}{\begin{minipage}{0.52\textwidth}\includegraphics[width=\textwidth]{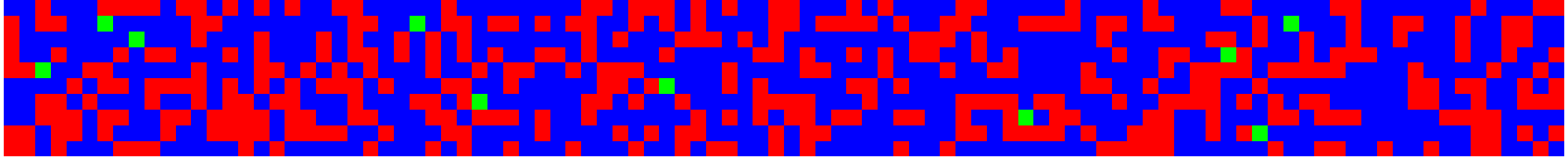}\end{minipage}}\\ 
	{\bf A:} All random, no siblings: 0.6\% Th0, 36.8\% Th1, 62.6\% Th2 
&
	{\bf B:} All random, with siblings: 1\% Th0, 38.4\% Th1, 60.6\% Th2\\

	\multicolumn{1}{c}{\begin{minipage}{0.52\textwidth}\includegraphics[width=\textwidth]{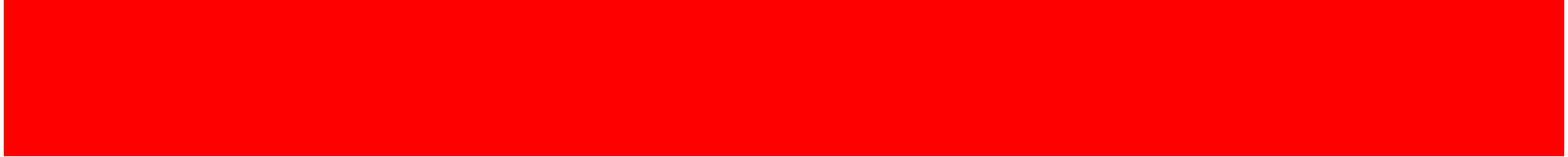}\end{minipage}} 
&
	\multicolumn{1}{c}{\begin{minipage}{0.52\textwidth}\includegraphics[width=\textwidth]{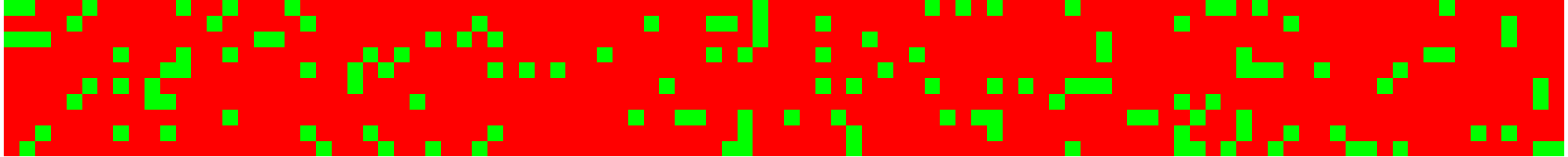}\end{minipage}}\\ 
	{\bf C:} $\mbox{GATA3}$: $100\%$ Th1
&
	{\bf D:} $\mbox{Tbet}$: $13.7\%$ Th0, $86.3\%$ Th1\\

	\multicolumn{1}{c}{\begin{minipage}{0.52\textwidth}\includegraphics[width=\textwidth]{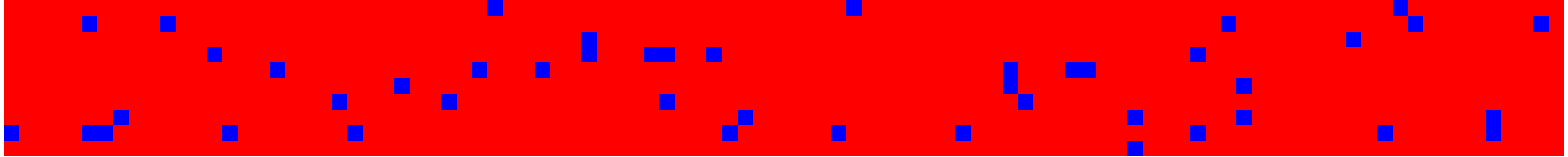}\end{minipage}} 
&
	\multicolumn{1}{c}{\begin{minipage}{0.52\textwidth}\includegraphics[width=\textwidth]{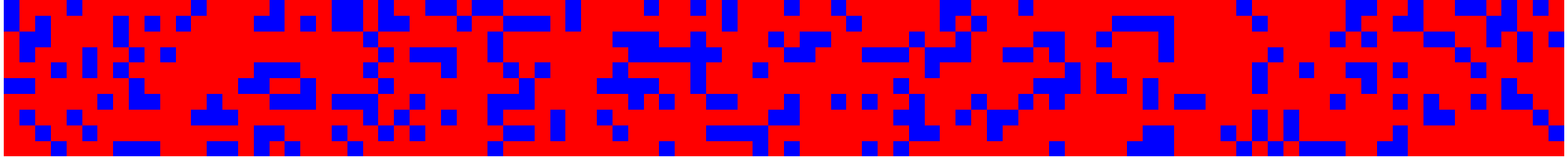}\end{minipage}}\\ 
	{\bf E:} $\mbox{IL4}$: $95.4\%$ Th1, $4.6\%$ Th2
&
	{\bf F:} $\mbox{IL4R}$: $73.3\%$ Th1, $26.7\%$ Th2\\


&\\
\end{tabular}}
\caption{\label{tableTh}
	Simulation results of the PBN showing the proportions of reached phenotypes. Green cells denote  Th0, red denote Th1 and blue denote Th1 phenotypes.
	Panel A) shows that the consideration of random functions in the set including the reference function and its (direct) parents for each components leads to the appearance of the three phenotypes;
	Panel B) shows that the results remain similar when including the siblings of the reference functions;
	Panel C) shows that, when random functions are considered only for GATA3, all the simulations lead to the sole Th1 phenotype;
	Panel D) shows that, when random functions are considered only for Tbet, most simulations lead to the Th1 phenotype, with a few simulations reverting to the Th0 phenotype;
	Panel E) shows that, when considering random functions for IL4, the Th0 phenotype does not show up, most simulations lead to Th1, with a few simulations leading to Th2;
	Panel F) shows that, when considering random functions for IL4R, the Th0 phenotype does not show up, most simulations lead to Th1, with simulations leading to Th2 in a higher proportion compared to Panel E.
}
\end{figure*}

 Starting from an initial state, in which all the components are inactive but IFNg, the simulations of the deterministic BN (synchronous, no probability associated to the functions) leads to a Th1 phenotype (with Tbet active). Using BoolNet,  1000 simulation runs are launched for the PBN defined as follows (see Figure \ref{tableTh} for the resulting proportions of reached phenotypes):
\begin{itemize}
 \item[A)] Associating random functions to each component: the reference function (with probability $\pi=0.8$) and its direct $p$ parents/children (each with probability $\pi=\frac{0.2}{p}$);
 \item[B)] Associating random functions to each component: the reference function (with probability $\pi=0.8$) and its direct $p$ parents/children and $s$ siblings (each with probability $\pi=\frac{0.2}{p+s}$);
\item[C)] Associating random functions to GATA3: the reference function with $\pi=0.8$ and its parents/children, each with $\pi=0.1$ ($p=2$);
\item[D)] Associating random functions to Tbet: the reference function with $\pi=0.8$ and its parents/children, each with $\pi=0.1$ ($p=2$);
\item[E)] Associating random functions to IL4: the reference function with $\pi=0.8$ and its parent $\pi=0.2$ (p=1);
\item[F)] Associating random functions to IL4R (the reference function with $\pi=0.8$ and its parent $\pi=0.2$ (p=1).
\end{itemize}

\section{\label{sec:conclusion}Conclusion and prospects}

The choice of appropriate functions to adequately reproduce desired dynamics is inherently hard due to the lack of regulatory data.
In this work, we have characterized the complexity of defining these functions in Boolean regulatory networks.
In particular, we have specified the Partial Ordered set (PO-Set) of the Boolean functions compatible with a given network topology.

Exploiting the {PO-Set} structure can be useful to tackle issues related to the definition and analysis of Boolean models. 
We have established a set of rules to compute the direct neighbors of any monotone Boolean function, without having to first generate the whole set of Boolean functions and subsequently compare them.
We have illustrated the usefulness of this procedure, which can be used to refine the definition of  random functions in probabilistic Boolean networks.

As a prospect, in problems related to model revision, the knowledge of the direct neighborhoods of regulatory functions would allow to perform local searches to improve model outcomes, with minimal impact on the regulatory structure.
Additionally, it would allow for the qualification of the set of models complying with certain requirements, such as: models that have the same regulatory graphs, but different functions; or models capable of satisfying similar dynamical restrictions.

Finally, although the proposed rules to uncover function neighbors apply to the case of Boolean functions, the extension to  multi-valued functions could be achieved through the Booleanization of the model \cite{Didier:2011aa}.

\subsubsection*{Availability}
The software implementing the rules to compute the parents and the children of a given Boolean function, is freely available at \url{https://github.com/ptgm/functionhood} under a GNU General Public License v3.0 (GPL-3.0). This software is expected to be made available as part of the set of software tools made available at \url{http://github.com/colomoto} by the \url{http://CoLoMoTo.org} (Consortium for Logical Models and Tools) consortium, and integrated into the GINsim modeling and simulation tool (\url{http://ginsim.org}).

\subsubsection*{Funding}
JC acknowledges the support from the Brazilian agency CAPES, with a one year research fellowship to visit IGC. This work has been further supported by the Portuguese national agency Fundação para a Ciência e a Tecnologia (FCT) with reference PTDC/EEI-CTP/2914/2014 (project ERGODiC) and UID/CEC/50021/2013.

\subsubsection*{Acknowledgments}
The authors thank Olga Zadvorna for her initial contribution to this work during her internship at IGC in 2015.

\bibliographystyle{plain}

\end{document}